\documentclass[pra,preprint,showpacs,preprintnumbers,amsmath,amssymb]{revtex4}
\usepackage{graphicx}% Include figure files
\usepackage{dcolumn}% Align table columns on decimal point
\usepackage{epic}%
\usepackage{eepic}%
\usepackage{bm}% bold math
\begin{document}

\preprint{APS/123-QED}

\newcommand{\scscs}{\scriptscriptstyle}

\title{Using the local density approximation and the LYP, BLYP, and B3LYP
functionals within Reference--State One--Particle Density--Matrix Theory}

\author{James P. Finley}

\affiliation{
Department of Physical Sciences,
Eastern New~Mexico University,
Station \#33, Portales, NM 88130}
\email{james.finley@enmu.edu}

\affiliation{Department of Applied Chemistry, Graduate School of Engineering,
The University of Tokyo, Tokyo, Japan 113-8656}

\date{\today}

\begin{abstract}
For closed-shell systems, the local density approximation (LDA) and the LYP, BLYP,
and B3LYP functionals are shown to be compatible with reference-state one-particle
density-matrix theory, where this recently introduced formalism is based on
Brueckner-orbital theory and an energy functional that includes exact exchange and
a non-universal correlation-energy functional. The method is demonstrated to
reduce to a density functional theory when the exchange-correlation
energy-functional has a simplified form, i.e., its integrand contains only the
coordinates of two electron, say $\mathbf{r}_1$ and $\mathbf{r}_2$, and it has a
Dirac delta function $\delta(\mathbf{r}_1-\mathbf{r}_2)$ as a factor.  Since
Brueckner and Hartree--Fock orbitals are often very similar, any local exchange
functional that works well with Hartree--Fock theory is a reasonable approximation
with reference-state one-particle density-matrix theory. The LDA approximation is
also a reasonable approximation. However, the Colle--Salvetti correlation-energy
functional, and the LYP variant, are not ideal for the method, since these are
universal functionals. Nevertheless, they appear to provide reasonable
approximations. The B3LYP functional is derived using a linear combination of two
functionals: One is the BLYP functional; the other uses exact exchange and a
correlation-energy functional from the LDA.

\end{abstract}

%\pacs{Valid PACS appear here}% PACS, the Physics and Astronomy Classification Scheme.
%\keywords{Suggested keywords}%Use showkeys class option if keyword display desired
\maketitle

\section{Introduction} \label{INTRO}

The Kohn-Sham version of density functional theory (DFT) plays a major role in
both quantum chemistry and condensed matter physics
\cite{Dreizler:90,Parr:89,Springborg:97,Ellis:95,Gross:94,Seminario:95,Handy:97}.
The local density approximation (LDA) \cite{Kohn:65} has been widely used for the
solid state, while for molecules, the most successful functional, a hybrid one
\cite{Becke:93,Burke:97,Perdew:96,Ernzerhof:96}, is known as B3LYP
\cite{Becke:93,Stephens:94}, where this functional contains three parameters, two
correlation-energy functionals, the Dirac-exchange functional with the a
correction, and exact exchange. The LYP correlation-energy functional
\cite{Lee:88} -- a key component of B3LYP -- is derived from the Colle--Salvetti
correlation-energy functional \cite{Colle:75}.

Recently, a density-matrix formalism was developed called reference-state
one-particle density-matrix theory \cite{Finley:bdmt.arxiv,Finley:bdmt}. This
method employs generalized Hartree--Fock equations containing the exact
exchange-potential and a nonlocal correlation-potential, where these equations are
obtained using the Brillouin-Brueckner condition (for non-variational energy
functionals) or functional minimization (for variational functionals).  The method
uses energy functionals that yield the target-state energy when the reference
state -- or its corresponding one-particle density-matrix -- is constructed from
Brueckner orbitals.  Unlike traditional DFT approaches, the $v$-representable
problem does {\em not} appear in the method, nor the need to introduce functionals
defined by a constrained search.  The correlation-energy functionals in the method
are non-universal, in the sense that they depend on the external
potential. Nevertheless, model systems can still be used to derive universal
energy-functionals. For example, the Colle--Salvetti functional was shown to be
compatible with the method as well as a one-particle-density-matrix variant of the
LDA. In addition, using time-independent many-body perturbation theory
\cite{Lindgren:86,Harris:92}, diagrammatic expansions were given for the
non-variational correlation-energy functionals that are expressed in terms of
orbitals and orbital energies.  When restrictions are placed on the orbital
energies, the individual diagrams are shown to {\em explicitly} depend on the
one-particle density-matrix of the reference state (and the external potential).

Below we demonstrate that for systems that can be treated with a simplified
exchange-correlation energy-functional -- leading to a local exchange-correlation
potential -- the reference-state one-particle density-matrix formalism reduces to
a density functional formalism, where the density that appears within the method is
the one from the reference state, and {\em not} the target state. This approach
differs in this way and in others from the formalism by Lindgren and Salomonson
\cite{Lindgren:02}, where their method is a generalization of the Hartree--Fock
Kohn--Sham approach \cite{Seidl:96} and yields orbitals that are believed to be
very similar to Brueckner orbitals.

The DFT functionals derived from a uniform electron gas are also shown to be
compatible with the reference-state one-particle density-matrix theory, where this
compatibility appears because the density of the target state for a uniform
electron gas is the same as the density of the reference state and, also, the
plane-wave states are Brueckner orbitals. In additional, functionals that
approximates the exact exchange-energy can also be used, since the exact
exchange-functional for Brueckner orbital theory, Hartree--Fock theory, and DFT
are the same. Furthermore, functionals derived using the one-particle
density-matrix of the reference state are compatible, assuming the errors due to
the external potential are small -- if the functional is a universal one.  Using
these observations, the LDA, and the BLYP and B3LYP functionals are shown to be
compatible with the reference-state one-particle density-matrix formalism. This
puts these functionals within an alternative DFT approach that avoids, for the
most part, existence theorems and constraint searches.

\section{Reference--State One--Particle Density--Matrix Theory} \label{dmt}

We seek an eigenfunction $|\Psi\rangle$ on the time-independent Hamiltonian
operator:
\begin{eqnarray} \label{H} %\mbox{\tiny H}
H = \sum_{ij} \left(
[i|({-}\mbox{\small$\frac{1}{2}$}\nabla^2)|j]
+ [i|v|j]\right)
a_i^\dagger a_j +  
\frac{1}{2}\sum_{ijkl} [ij|kl] 
a_i^\dagger a_k^\dagger a_l a_j,
\end{eqnarray} 
where the integrals are written using chemist's notation \cite{Szabo:82}: 
\begin{eqnarray} \label{chemist} %\mbox{\tiny chemist} 
\mbox{} [ij|kl]&=&\sum_{\omega_2\omega_2} \int 
\psi_{i}^*(\mathbf{x}_1)
\psi_{j}(\mathbf{x}_1) 
r_{12}^{-1}
\psi_{k}^*(\mathbf{x}_2) 
%\psi_{l}(\mathbf{x}_2) \; d \mathbf{r}_1 d \mathbf{r}_{2.} \;\;\;\;\;
\psi_{l}(\mathbf{x}_2) \; d \mathbf{r}_1 d \mathbf{r}_{2},
\end{eqnarray} 
and the spatial and spin coordinates, $\mathbf{r}$ and $\omega$, are denoted
collectively by $\mathbf{x}$.  The eigenfunction of interest, or target state
$|\Psi\rangle$, yields the electronic energy ${\cal E}$:
\begin{eqnarray} \label{E} %\mbox{\tiny E}
{\cal E} = \frac{\langle\Psi|H|\Psi\rangle}{\langle\Psi|\Psi\rangle}.
\end{eqnarray} 

Consider a single-determinantal reference-state, say $|\Phi\rangle$, where this
state is a first-order approximation of the target state; the first-order energy
is
\begin{equation} \label{E1} %\mbox{\tiny E1}
E_1[\Phi] = \langle \Phi|H|\Phi\rangle. 
\end{equation}
Adding and subtracting $E_1[\Phi]$ on the left side of Eq.~(\ref{E}), gives
\begin{eqnarray} \label{Eb} %\mbox{\tiny Eb}
{\cal E} = E_1 + ({\cal E} - E_1) = E_1[\Phi] + {\cal E}_{\mathrm{co}}[\Phi],
\end{eqnarray} 
where this Eq.\ defines the correlation energy ${\cal E}_{\mathrm{co}}[\Phi]$
as (${\cal E} - E_1$).

There is a one-to-one correspondence between determinant states and their
one-particle density-matrices \cite{Blaizot:86,Parr:89}, where these
density-matrices $\gamma$ are given
by~\cite{Dirac:30,Dirac:31,Lowdin:55a,Lowdin:55b,McWeeny:60}
\begin{equation} \label{dm} %\mbox{\tiny dm}
\gamma(\mathbf{x},\mathbf{x^\prime}) 
= \sum_{w\in \{\psi_o\rightarrow \Phi\}} 
\psi_{w}(\mathbf{x}) \psi_{w}^*(\mathbf{x^\prime});
\;\; \gamma \longleftrightarrow |\Phi\rangle, 
\end{equation} 
and the notation on the right-hand-side indicates the one-to-one correspondence;
furthermore, the sum over $w$ includes only the {\it o}ccupied orbitals of the
reference state $|\Phi\rangle$: This orbital set is denoted by
$\{\psi_o\rightarrow \Phi\}$; the {\it u}noccupied set is denoted by
$\{\psi_u\rightarrow \Phi\}$. Unless stated otherwise, two sets of occupied or
unoccupied orbitals that differ by a unitary transformation are considered
equivalent.

Because of the one-to-one correspondence mentioned above, any functional that
depends on the reference state $|\Phi\rangle$ can be transformed into one that
depends on the one-particle density-matrix $\gamma$. For example, by requiring the
external potential $v(\mathbf{r})$ to be a spin-free operator, the first-order
energy can be written as
\begin{eqnarray} \label{first.eb} %\mbox{\tiny first.eb}
E_1[\gamma]
=
\int \left[{-}\mbox{\small$\frac{1}{2}$}\nabla_{\mathbf{r}}^2\,
\gamma(\mathbf{x},\mathbf{x^\prime}) 
\right]_{\mbox{\tiny $\mathbf{x^\prime}\!\!=\!\!\mathbf{x}$}}
\!d\mathbf{x} + 
\int v(\mathbf{r}) 
\gamma(\mathbf{x},\mathbf{x})
\,d\mathbf{x} 
+ E_{\mathrm{J}}[\gamma] + E_{\mathrm{x}}[\gamma], 
\end{eqnarray}
where the Coulomb and exchange energies have their usual forms:
\begin{eqnarray} \label{coul.dm} %\mbox{\tiny coul.dm}
E_{\mathrm{J}}[\gamma]&=&
\frac12 \int \!\! \int 
r_{12}^{-1}
\gamma(\mathbf{x_1},\mathbf{x_1}) 
\gamma(\mathbf{x_2},\mathbf{x_2})
\,d\mathbf{x_1}
\,d\mathbf{x_2},
\\ \label{exch.dm} %\mbox{\tiny exch.dm}
\mbox{$-E_{\mathrm{x}}[\gamma]$} &=&
\frac12 \int\!\!\int 
r_{12}^{-1}
\gamma(\mathbf{x_1},\mathbf{x_2}) 
\gamma(\mathbf{x_2},\mathbf{x_1})
\,d\mathbf{x_1}
\,d\mathbf{x_2},
\end{eqnarray}
and, henceforth, an integration over $\mathbf{x}$ implies an integration over
$\mathbf{r}$ and a summation over~$\omega$.

\subsection{Hartree--Fock Theory} \label{HFT} %\mbox{\tiny HFT}

Let the Fock operator be denoted $\hat{F}_\gamma$; the functional derivative of
the first-order energy $E_1[\gamma]$ gives the kernel of this operator
\cite{Parr:89}:
\begin{eqnarray} \label{kernfock} %\mbox{\tiny kernfock}
F_\gamma(\mathbf{x}_1,\mathbf{x}_2)
&=&
\frac{\delta E_1[\gamma]} 
{\delta \gamma(\mathbf{x}_2,\mathbf{x}_1)} 
%\hspace{58ex}
\\ \nonumber 
%\hspace{20ex}
&=&
\delta(\mathbf{x}_2-\mathbf{x}_1) \left(
-\mbox{\small$\frac{1}{2}$}
\nabla_{\mbox{\tiny $2$}}^2
+ v(\mathbf{r}_2)
+
\int 
r_{23}^{-1}
\gamma(\mathbf{x}_3,\mathbf{x}_3) 
\,d\mathbf{x}_3
\right)
+
v_{\mathrm{x}}^{\scscs \gamma}(\mathbf{x}_1,\mathbf{x}_2),
\end{eqnarray}
where the two-body function, $v_{\mathrm{x}}^{\scscs \gamma}(\mathbf{x}_1,\mathbf{x}_2)$,
is obtained from the exchange energy:
\begin{eqnarray} \label{kernexch} %\mbox{\tiny kernexch}
v_{\mathrm{x}}^{\scscs \gamma}(\mathbf{x}_1,\mathbf{x}_2)
=
\frac{\delta E_{\mathrm{x}}[\gamma]} 
{\delta \gamma(\mathbf{x}_2,\mathbf{x}_1)} 
=
-r_{12}^{-1}
\gamma(\mathbf{x}_1,\mathbf{x}_2),
\end{eqnarray}
and this function is the kernel of the exchange operator,
denoted by $\hat{v}_{\mathrm{x}}^{\scscs \gamma}$.

Minimizing the functional $E_1[\gamma]$, subject to the constraint that the
one-particle density-matrix comes from a single-determinantal state, yields
$E_1[\tilde{\tau}]$, where $\tilde{\tau}$ is the one-particle density-matrix of
the Hartree--Fock reference-state $|\tilde{\Theta}\rangle$:
\begin{equation} \label{hfdm} %\mbox{\tiny hfdm}
\tilde{\tau}(\mathbf{x},\mathbf{x^\prime}) 
= \sum_{w\in \{\psi_o\rightarrow \tilde{\Theta}\}} 
\psi_{w}(\mathbf{x}) \psi_{w}^*(\mathbf{x^\prime}),
\end{equation} 
and this density-matrix satisfies the following
condition:
\begin{subequations}
\label{F.noncan} %\mbox{\tiny F.noncan}
\begin{eqnarray} \label{F.noncana} %\mbox{\tiny F.noncana}
\langle \psi_r
|\hat{F}_{\tilde{\tau}}|
\psi_w\rangle=0; \;\; \psi_w\in \{\psi_o\rightarrow \tilde{\tau}\}, \;\;
\psi_r\in \{\psi_u\rightarrow \tilde{\tau}\},
\end{eqnarray}
and the right-hand-side notation indicates that $\psi_w$ is a member of the set of
occupied orbital that determines $\tilde{\tau}$ (and $|\tilde{\Theta}\rangle$);
$\psi_r$ is unoccupied.  The operator form of this Eq.\ is
\begin{eqnarray} \label{F.noncanb} %\mbox{\tiny F.noncanb}
\left(\mbox{\small $\hat{1}$}-\tilde{\tau}\right)\hat{F}_{\tilde{\tau}} \tilde{\tau} =0,
\end{eqnarray}
\end{subequations}
where the identity operator $\hat{1}$, using a Hartree--Fock basis set, is given by
\begin{eqnarray} \label{ident.op} %\mbox{\tiny ident.op}
\hat{1}\;=\sum_{w\in \{\psi_o\rightarrow \tilde{\tau}\}} 
|\psi_{w}\rangle\langle\psi_{w}|
\;\; +\!
\sum_{r\in \{\psi_u\rightarrow \tilde{\tau}\}} 
|\psi_{r}\rangle\langle\psi_{r}|.
\end{eqnarray}

A unique set of occupied and unoccupied orbitals is obtained by requiring the
occupied and unoccupied blocks of $\hat{F}_{\tilde{\tau}}$ to be diagonal:
\begin{subequations}
\label{F.can} %\mbox{\tiny F.can}
\begin{eqnarray} 
\hat{F}_{\tilde{\tau}}
\psi_w^{\tilde{\tau}}(\mathbf{x})&=&
\varepsilon_{w}^{\mbox{\tiny $\tilde{\tau}$}}
\psi^{\tilde{\tau}}_w(\mathbf{x}),
\;\; \psi^{\tilde{\tau}}_w\in \{\psi_o\rightarrow \tilde{\tau}\},
\\
\hat{F}_{\tilde{\tau}}
\psi^{\tilde{\tau}}_r(\mathbf{x})&=&
\varepsilon_{r}^{\mbox{\tiny $\tilde{\tau}$}}
\psi^{\tilde{\tau}}_r(\mathbf{x}), \;\;
\psi^{\tilde{\tau}}_r\in \{\psi_u\rightarrow \tilde{\tau}\}.
\end{eqnarray}
\end{subequations}
The orbitals sets that satisfy Eqs.~(\ref{F.noncan}) and (\ref{F.can}) are denoted
by $\{\psi^{\tilde{\tau}}_o \leftarrow \tilde{\tau}, \mbox{\small
$\hat{F}_{\tilde{\tau}}$}\}$ and
\mbox{$\{\psi^{\tilde{\tau}}_u\leftarrow\tilde{\tau},\mbox{\small
$\hat{F}_{\tilde{\tau}}$}\}$}, indicating that they are uniquely determined by
$\tilde{\tau}$ and $\hat{F}_{\tilde{\tau}}$.

Substituting Eq.~(\ref{kernfock}) into Eqs.~(\ref{F.can}) gives the canonical
 Hartree--Fock Eqs:
\begin{eqnarray} \label{F.canb} %\mbox{\tiny F.canb}
\left(
-\mbox{\small$\frac{1}{2}$}
\nabla_{\mbox{\tiny $1$}}^2
+ v(\mathbf{r}_1)
+
\int 
r_{12}^{-1}
\tilde{\tau}(\mathbf{x}_2,\mathbf{x}_2) 
\,d\mathbf{x}_2
+ \hat{v}_{\mathrm{x}}^{\scscs \tilde{\tau}}(\mathbf{x}_1)
\right) 
\psi^{\tilde{\tau}}_i(\mathbf{x}_1)
=
\varepsilon_{i}^{\mbox{\tiny $\tilde{\tau}$}}
\psi^{\tilde{\tau}}_i(\mathbf{x}_1).
\end{eqnarray}

\subsection{Variational Brueckner orbital theory} \label{VBOT} %\mbox{\tiny VBOT}

Brueckner orbital theory
\cite{Brueckner:54,Nesbet:58,Brenig:61,Lowdin:62,Kutzelnigg:64,Cizek:80,Chiles:81,Stolarczyk:84,Handy:85,Handy:89,Raghavachari:90,Hirao:90,Stanton:92,Hampel:92,Scuseria:94,Lindgren:02}
is a generalization of Hartree--Fock theory that utilizes a single-determinantal
state that has the maximum overlap with the target state
\cite{Kobe:71,Shafer:71}.  We now review a variant of Brueckner orbital theory
that is used to derive the reference-state one-particle density-matrix theory
\cite{Finley:bdmt}.

In this previous work, we introduced four trial wavefunctions -- say
$|\Psi_\Phi^{\scscs(\eta)}\rangle$, where ($\eta= \mbox{{\small I, II, III}, and
{\small IV}}$) -- that are defined with respect to a single-determinantal
reference-state $|\Phi\rangle$.  The first trial-wavefunction $|\Psi_\Phi^{\scscs
(\mathrm{I})}\rangle$ is simply the target state of interest, $|\Psi\rangle$, with
the single excitations removed: 
\begin{equation} \label{} %\mbox{}
|\Psi_\Phi^{\scscs (\mathrm{I})}\rangle = 
\left(1-P_{11}^{\Phi}\right)|\Psi\rangle,
\end{equation}
where the projector for the singly-excited states is
\begin{equation} \label{P11} \mbox{\tiny P11}
P_{11}^{\Phi} = 
\sum_{w\in \{\psi_o\rightarrow \Phi\}} 
\sum_{r\in \{\psi_u\rightarrow \Phi\}}
|\Phi_{w}^{r}\rangle\langle\Phi_{w}^{r}|,
%\;\; $P11$
\end{equation}
and the singly-excited state are
\begin{equation}
|\Phi_{w}^{r}\rangle= a^\dagger_{r} a_{w} |\Phi\rangle; \; \;
\psi_w \in \{\psi_o\rightarrow \Phi\}, \; \; \psi_r \in \{\psi_u\rightarrow \Phi\}. 
\end{equation}
The $P_{11}^{\Phi}$ subspace is completely determined by $|\Phi\rangle$;
$P_{11}^{\Phi}$ is also invariant to a unitary transformation of occupied, or
virtual, orbitals \cite{Stolarczyk:84}.

The second trial-wavefunction $|\Psi_\Phi^{\scscs (\mathrm{II})}\rangle$ is
defined with respect to the target state expressed by an exponential ansatz:
($|\Psi\rangle= e^{S_{\Phi}}|\Phi\rangle$), where $|\Psi_\Phi^{\scscs
(\mathrm{II})}\rangle$ is generated by removing the single-excitation amplitudes
$S_1^{\Phi}$ from the cluster-operator $S$:
\begin{equation} \label{trial.IIb} %\mbox{\tiny trial.IIb}
|\Psi_\Phi^{\scscs (\mathrm{II})}\rangle=e^{(S_{\Phi}-S_1^{\Phi})}|\Phi\rangle.
\end{equation}

The third trial-wavefunction $|\Psi_\Phi^{\scscs (\mathrm{III})}\rangle$ can be
generated by its wave-operator:
\begin{equation} \label{ccf.wo} %\mbox{\tiny ccf.wo}
\hat{\Omega}_\Phi|\Phi\rangle = |\Psi_\Phi^{\scscs (\mathrm{III})}\rangle,
\end{equation}
that can be expressed in an exponential form: ($\hat{\Omega}_\Phi=
e^{\hat{S}_{\Phi}}|\Phi\rangle$), where $\hat{S}_{\Phi}$ can be written as a sum
$n$-body excitations, with the exclusion of a one-body operator:
\begin{equation} \label{T.ccf} %\mbox{\tiny T.ccf}
\hat{S}_{\Phi}= \hat{S}_2^{\Phi}+\hat{S}_3^{\Phi}+\cdots.
\end{equation}
The wave operator $\hat{\Omega}_{\Phi}$ is as a solution to the coupled cluster
equations
\cite{Hubbard:57,Coester:58,Cizek:66,Cizek:69,Cizek:71,Lindgren:78,Bartlett:78,
Pople:78,Lindgren:86,Harris:92} with the single excitation portion removed:
\begin{equation} \label{lct.op.iv} %\mbox{\tiny lct.op.iv}
\left(1-P_{11}^{\Phi}\right) \left(H\hat{\Omega}_\Phi \right)_{\text{op,cn}} = 0,
\end{equation}
where only the open (op) and connected (cn) portions enter into the relation. This
expression defines the trial functional $|\Psi_\Phi^{\scscs
(\mathrm{III})}\rangle$ using Eq.~(\ref{ccf.wo}).

The fourth trial wavefunctions $|\Psi_\Phi^{\scscs (\mathrm{IV})}\rangle$ is not
considered here, except to mention that it is obtained by solving the
configuration-interaction equations \cite{Schaefer:72,Szabo:82,Harris:92} in an
approximate way, i.e., by neglecting the single-excitation portion.

All of the trial states $|\Psi_\Phi^{\scscs (\eta)}\rangle$ share the property
that they contain no single excitations. i.e., ($P_{11}^{\Phi}|\Psi_\Phi^{\scscs
(\eta)}\rangle = 0$), and they generate the target state $|\Psi\rangle$ when their
reference state satisfies ($|\Phi\rangle=|\Theta\rangle$), where $|\Theta\rangle$
is the determinantal state constructed from occupied Bruckner orbitals. In other
words, we have
\begin{equation} \label{trwf=exact} %\mbox{\tiny trwf=exact}
|\Psi_\Theta^{\scscs (\eta)}\rangle = |\Psi\rangle.
\end{equation}

Because of the one-to-one correspondence mentioned above, the trial wavefunctions
depend on the one-particle density-matrix; so we can write $|\Psi_\gamma^{\scscs
(\eta)}\rangle$. Using this notation, we can define variational energy-functionals
that depend on the one-particle density-matrix:
\begin{eqnarray} \label{Efuncts.var} %\mbox{\tiny Efuncts.var}
\bar{E}_\eta[\gamma] = 
\frac{\langle\Psi_\gamma^{\scscs (\eta)}|
H|\Psi_\gamma^{\scscs (\eta)}\rangle}
{\langle\Psi_\gamma^{\scscs (\eta)}|\Psi_\gamma^{\scscs (\eta)}\rangle}
= E_1[\gamma] + \bar{E}_{\mathrm{co}}^{\scscs (\eta)}[\gamma],
\end{eqnarray}
where the last relation defines the correlation-energy functionals
$\bar{E}_{\mathrm{co}}^{\scscs (\eta)}[\gamma]$ as
($\bar{E}_\eta[\gamma]-E_1[\gamma]$).

Substituting Eq.~(\ref{first.eb}) into Eqs.~(\ref{Eb}) and (\ref{Efuncts.var}) gives
the following:
\begin{subequations} \label{E.Ef} %\mbox{\tiny E.Ef}
\begin{eqnarray} 
{\cal E}
&=&
\int \left[{-}\mbox{\small$\frac{1}{2}$}\nabla_{\mathbf{r}}^2\,
\gamma(\mathbf{x},\mathbf{x^\prime}) 
\right]_{\mbox{\tiny $\mathbf{x^\prime}\!\!=\!\!\mathbf{x}$}}
\!d\mathbf{x} + 
\int v(\mathbf{r}) 
\gamma(\mathbf{x},\mathbf{x})
\,d\mathbf{x} 
+ E_{\mathrm{J}}[\gamma] + {\cal E}_{\mathrm{xc}}[\gamma], 
\\
\bar{E}_\eta[\gamma]
&=&
\int \left[{-}\mbox{\small$\frac{1}{2}$}\nabla_{\mathbf{r}}^2\,
\gamma(\mathbf{x},\mathbf{x^\prime}) 
\right]_{\mbox{\tiny $\mathbf{x^\prime}\!\!=\!\!\mathbf{x}$}}
\!d\mathbf{x} + 
\int v(\mathbf{r}) 
\gamma(\mathbf{x},\mathbf{x})
\,d\mathbf{x} 
+ E_{\mathrm{J}}[\gamma] + 
\bar{E}_{\mathrm{xc}}^{\scscs (\eta)}[\gamma],
\end{eqnarray}
\end{subequations}
where the exchange-correlation energy and exchange-correlation energy-functionals
are, respectively, defined by
\begin{subequations} \label{Ex.Exf} %\mbox{\tiny Ex.Exf}
\begin{eqnarray} 
{\cal E}_{\mathrm{xc}}[\gamma]
= 
E_{\mathrm{x}}[\gamma] + {\cal E}_{\mathrm{co}}[\gamma],\\
\label{Excorr.v} %\mbox{\tiny Excorr.v}
\bar{E}_{\mathrm{xc}}^{\scscs (\eta)}[\gamma] = 
E_{\mathrm{x}}[\gamma]
+
\bar{E}_{\mathrm{co}}^{\scscs (\eta)}[\gamma] .
\end{eqnarray}
\end{subequations}

We now generalize many of the Hartree--Fock relations from the previous subsection
simply by replacing $E_1[\gamma]$ with $\bar{E}_\eta[\gamma]$ and the Fock
operator $\hat{F}_\gamma$ with generalized, or exact, Fock operators, ${\cal
\hat{\zeta}}_{\gamma}^{\scscs (\eta)}$.

The functional derivative of our energy functionals yield two-body functions that
serve as the kernels of the exact Fock operators:
\begin{eqnarray} \label{kernEfock} %\mbox{\tiny kernEfock}
{\cal \zeta}_{\gamma}^{\scscs (\eta)}(\mathbf{x}_1,\mathbf{x}_2)&=&
\frac{\bar{E}_\eta[\gamma]}
{\delta \gamma(\mathbf{x}_2,\mathbf{x}_1)} 
%\hspace{58ex}
\\ \nonumber 
%\hspace{20ex}
&=&
\delta(\mathbf{x}_2-\mathbf{x}_1) \left(
-\mbox{\small$\frac{1}{2}$}
\nabla_{\mbox{\tiny $2$}}^2
+ v(\mathbf{r}_2)
+
\int 
r_{23}^{-1}
\gamma(\mathbf{x}_3,\mathbf{x}_3) 
\,d\mathbf{x}_3
\right)
+
\nu_{\mathrm{xc}}^{\scscs \gamma\eta}(\mathbf{x}_1,\mathbf{x}_2),
\end{eqnarray}
where the kernels of the exchange-correlation operators,
$\nu_{\mathrm{xc}}^{\scscs \gamma\eta}(\mathbf{x}_1,\mathbf{x}_2)$, are obtained
from the exchange-correlation energy-functionals:
\begin{eqnarray} \label{kernexchcorr} %\mbox{\tiny kernexchcorr}
\nu_{\mathrm{xc}}^{\scscs \gamma\eta}(\mathbf{x}_1,\mathbf{x}_2)
=
\frac{\delta \bar{E}_{\mathrm{xc}}^{\scscs (\eta)}[\gamma]} 
{\delta \gamma(\mathbf{x}_2,\mathbf{x}_1)} 
=
\frac{\delta \bar{E}_{\mathrm{co}}^{\scscs (\eta)}[\gamma]} 
{\delta \gamma(\mathbf{x}_2,\mathbf{x}_1)}
-r_{12}^{-1}
\gamma(\mathbf{x}_1,\mathbf{x}_2),
\end{eqnarray}
where the last relation uses Eqs.~(\ref{Excorr.v}) and (\ref{kernexch}).

We now require the target state to be a ground state; using the variation theorem
and Eq.~(\ref{trwf=exact}), the minimizing of the functionals
$\bar{E}_\eta[\gamma]$ -- subject to the constraint that the one-particle
density-matrix comes from a single-determinantal state -- yields
\begin{eqnarray} \label{trf=exact} %\mbox{\tiny trf=exact}
{\cal E}=\bar{E}_\eta[\tau],
\end{eqnarray}
and from Eqs.~(\ref{E.Ef}) and (\ref{Ex.Exf}) we have
\begin{eqnarray} \label{co=cof} %\mbox{\tiny co=cof}
{\cal E}_{\mathrm{co}}[\tau]&=&\bar{E}_{\mathrm{co}}^{\scscs (\eta)}[\tau], \\
{\cal E}_{\mathrm{xc}}[\tau]&=&\bar{E}_{\mathrm{xc}}^{\scscs (\eta)}[\tau],
\end{eqnarray}
where $\tau$ is the one-particle density-matrix of the Brueckner reference-state
$|\Theta\rangle$:
\begin{equation} \label{bdm} %\mbox{\tiny bdm}
\tau(\mathbf{x},\mathbf{x^\prime}) 
= \sum_{w\in \{\psi_o\rightarrow \Theta\}} 
\psi_{w}(\mathbf{x}) \psi_{w}^*(\mathbf{x^\prime}),
\end{equation} 
and the Brueckner orbitals satisfy the following equivalent conditions:
\begin{subequations}
\label{eF.noncan} %\mbox{\tiny eF.noncan}
\begin{eqnarray} \label{eF.noncana} %\mbox{\tiny eF.noncana}
\langle \psi_r
|{\cal \hat{\zeta}}_{\tau}^{\scscs (\eta)}|
\psi_w\rangle&=&0; \;\; \psi_w\in \{\psi_o\rightarrow \tau\}, \;\;
\psi_r\in \{\psi_u\rightarrow \tau\}, \\
\label{eF.noncanbX} %\mbox{\tiny eF.noncanb}
\left(\mbox{\small $\hat{1}$}-\tau\right)
{\cal \hat{\zeta}}_{\tau}^{\scscs (\eta)}
\tau&=&0,
\end{eqnarray}
\end{subequations}
where these orbitals do not depend of $\eta$ -- any trial wavefunction gives the
same results.

A unique set of occupied and unoccupied orbitals is obtained by requiring the
occupied and unoccupied blocks of ${\cal \hat{\zeta}}_{\tau}^{\scscs (\eta)}$ to
be diagonal:
\begin{subequations}
\label{eF.can} %\mbox{\tiny eF.can}
\begin{eqnarray} 
{\cal \hat{\zeta}}_{\tau}^{\scscs (\eta)}
\psi_w^{\tau}(\mathbf{x})&=&
\xi_{w}^{\mbox{\tiny $\tau$}}
\psi^{\tau}_w(\mathbf{x}),
\;\; \psi^{\tau}_w\in \{\psi_o\rightarrow \tau\},
\\
{\cal \hat{\zeta}}_{\tau}^{\scscs (\eta)}
\psi^{\tau}_r(\mathbf{x})&=&
\xi_{r}^{\mbox{\tiny $\tau$}}
\psi^{\tau}_r(\mathbf{x}), \;\;
\psi^{\tau}_r\in \{\psi_u\rightarrow \tau\}.
\end{eqnarray}
\end{subequations}
Henceforth, the orbitals sets that satisfy Eqs.~(\ref{eF.noncan}) and
(\ref{eF.can}) are denoted by $\{\psi^{\tau}_o \leftarrow \tau, \mbox{\small
$\hat{\zeta}_{\tau}^{\scscs (\eta)}$}\}$ and $\{\psi^{\tau}_u \leftarrow \tau,
\mbox{\small $\hat{\zeta}_{\tau}^{\scscs (\eta)}$}\}$, indicating that they are
determined by $\tau$ and ${\cal \hat{\zeta}}_{\tau}^{\scscs (\eta)}$.  Since
theses orbitals, and their energies, can, perhaps, depend on $\eta$, it is more
precise to denote then by $\psi^{\tau\eta}_i$ and $\xi_{i}^{\mbox{\tiny
$\tau\eta$}}$, but we suppress the $\eta$ superscripts to keep the notation less
cluttered.

Substituting Eq.~(\ref{kernEfock}) into Eqs.~(\ref{eF.can}) gives generalized,
canonical Hartree--Fock Eqs:
\begin{eqnarray} \label{eF.canb} %\mbox{\tiny eF.canb}
\left(
-\mbox{\small$\frac{1}{2}$}
\nabla_{\mbox{\tiny $1$}}^2
+ v(\mathbf{r}_1)
+
\int 
r_{12}^{-1}
\tau(\mathbf{x}_2,\mathbf{x}_2) 
\,d\mathbf{x}_2
+ \hat{\nu}_{\mathrm{xc}}^{\scscs \tau\eta}(\mathbf{x}_1)
\right) 
\psi^{\tau}_i(\mathbf{x}_1)
=
\xi_{i}^{\mbox{\tiny $\tau$}}
\psi^{\tau}_i(\mathbf{x}_1).
\end{eqnarray}

\section{Closed-Shell Case}

Our interest here is in cases where the Hamiltonian is spin-free -- it contains no
spin coordinates -- and the target state $|\Psi\rangle$ is a ground-state singlet
that is well described by a closed-shell reference-state $|\Phi\rangle$.  In these
cases -- and many others -- it is appropriate to use a set of spatially restricted
orbitals, given by
\begin{equation} \label{spinorb.res} %\mbox{\tiny spinorb.res}
\psi_{j\sigma}(\mathbf{x})=
\chi_j(\bm{r})\sigma(\omega); \;\; \sigma=\alpha,\beta,
\end{equation}
so that $|\Phi\rangle$ is determined by a set of doubly-occupied spatial orbitals,
denoted by $\{\chi_o\mbox{\footnotesize $\rightarrow\Phi$}\}$, where this set also
determines the virtual set, denoted by $\{\chi_u\mbox{\footnotesize
$\rightarrow\Phi$}\}$. Two sets of orbitals that differ by a unitary
transformation are, again, considered equivalent.

By definition, the spinless, one-particle density-matrix is given by \cite{McWeeny:60,Parr:89}
\begin{equation}
\rho_1(\mathbf{r}_1,\mathbf{r}_2)=
\sum_\omega \gamma(\mathbf{r}_1,\omega;\mathbf{r_2},\omega),
\end{equation}
and from Eqs.\ (\ref{spinorb.res}) and (\ref{dm}) with ($w=x\sigma$) we have
\begin{equation}
\rho_1(\mathbf{r}_1,\mathbf{r}_2)=
\;\;2\!\!\!\!\!\!\sum_{x\in \{\chi_o\rightarrow \Phi\}}
\chi_{x}(\mathbf{r}_1) \chi_{x}^*(\mathbf{r}_2);
\;\; \rho_1 \longleftrightarrow |\Phi\rangle.
\end{equation}
By using the restricted orbitals defined by Eq.~(\ref{spinorb.res}), it is easily
demonstrated that the one-particle density-matrix $\gamma$ is determined by the
spinless one, as indicated by the following relation:
\begin{equation} \label{gamma.sf} %\mbox{\tiny gamma.sf}
\gamma(\mathbf{x}_1,\mathbf{x}_2) = \frac12 \rho_1(\mathbf{r}_1,\mathbf{r}_2)
\delta_{\omega_1\omega_2}.
\end{equation}
Hence, any functional of $\gamma$ now becomes a functional of $\rho_1$.  In
particular, for a spin-free Hamiltonian, the spin-variable summations are
easily performed, yielding explicit functionals of $\rho_1$.  For example, the
electronic energy and our energy-functionals, Eqs.~(\ref{Eb}) and
(\ref{Efuncts.var}), can be written as
\begin{subequations}
\label{Efuncts.sf} %\mbox{\tiny Efuncts.sf}
\begin{eqnarray} 
\label{Eb.sf} %\mbox{\tiny Eb.sf}
{\cal E}&=&E_1[\rho_1] + {\cal E}_{\mathrm{co}}[\rho_1],\\
\label{Efuncts.var.sf} %\mbox{\tiny Efuncts.var.sf}
\bar{E}_\eta[\rho_1] 
&=&E_1[\rho_1] + \bar{E}_{\mathrm{co}}^{\scscs (\eta)}[\rho_1], 
\end{eqnarray}
\end{subequations}
or the representations of these functionals, given by Eqs.~(\ref{E.Ef}), reduce
to
\begin{subequations} 
\label{Efuncts.sfb} %\mbox{\tiny Efuncts.sfb}
\begin{eqnarray} 
{\cal E}
=
\int \left[{-}\mbox{\small$\frac{1}{2}$}\nabla_{\mathbf{r}}^2\,
\rho_1(\mathbf{r},\mathbf{r}^\prime) 
\right]_{\mbox{\tiny $\mathbf{r^\prime}\!\!=\!\!\mathbf{r}$}}
\!d\mathbf{r} + 
\int v(\mbox{$\mathbf{r}$}) 
\rho(\mathbf{r}) 
\,d\mathbf{r}
+ E_{\mathrm{J}}[\rho] +
{\cal E}_{\mathrm{xc}}[\rho_1], \\
\bar{E}_\eta[\rho_1] 
=
\int \left[{-}\mbox{\small$\frac{1}{2}$}\nabla_{\mathbf{r}}^2\,
\rho_1(\mathbf{r},\mathbf{r}^\prime) 
\right]_{\mbox{\tiny $\mathbf{r^\prime}\!\!=\!\!\mathbf{r}$}}
\!d\mathbf{r} + 
\int v(\mbox{$\mathbf{r}$}) 
\rho(\mathbf{r}) 
\,d\mathbf{r}
+ E_{\mathrm{J}}[\rho] +
\bar{E}_{\mathrm{xc}}^{\scscs (\eta)}[\rho_1],
\end{eqnarray}
\end{subequations} 
where the definitions given by Eqs.~(\ref{Ex.Exf}) can be expressed using a
dependence upon $\rho_1$:
\begin{subequations} 
\label{exco.sf} %\mbox{\tiny exco.sf}
\begin{eqnarray} 
{\cal E}_{\mathrm{xc}}[\rho_1]=
E_{\mathrm{x}}[\rho_1] + 
{\cal E}_{\mathrm{co}}[\rho_1],\\
\label{exco.sfv} %\mbox{\tiny exco.sfv}
\bar{E}_{\mathrm{xc}}^{\scscs (\eta)}[\rho_1]=
E_{\mathrm{x}}[\rho_1]
+ 
\bar{E}_{\mathrm{co}}^{\scscs (\eta)}[\rho_1];
\end{eqnarray}
\end{subequations} 
furthermore, the first-order energy, Eq.\ (\ref{first.eb}), becomes
\begin{eqnarray} \label{first.eb.sf} %\mbox{\tiny first.eb.sf} 
E_1[\rho_1]
=
\int \left[{-}\mbox{\small$\frac{1}{2}$}\nabla_{\mathbf{r}}^2\,
\rho_1(\mathbf{r},\mathbf{r}^\prime) 
\right]_{\mbox{\tiny $\mathbf{r^\prime}\!\!=\!\!\mathbf{r}$}}
\!d\mathbf{r} + 
\int v(\mbox{$\mathbf{r}$}) 
\rho(\mathbf{r}) 
\,d\mathbf{r}
+ E_{\mathrm{J}}[\rho] + E_{\mathrm{x}}[\rho_1],
\end{eqnarray}
where the Coulomb and exchange energies, Eq.\ (\ref{coul.dm}) and (\ref{exch.dm}),
reduce to
\begin{eqnarray} \label{coul.dm.sf} %\mbox{\tiny coul.dm.sf}
E_{\mathrm{J}}[\rho]&=&
\frac12 \int \!\! \int 
r_{12}^{-1}
\rho(\mathbf{r}_1) 
\rho(\mathbf{r}_2) 
\,d\mathbf{r}_1
\,d\mathbf{r}_2,
\\ \label{exch.dm.sf} %\mbox{\tiny exch.dm.sf}
\mbox{$-E_{\mathrm{x}}[\rho_1]$} &=&
\frac14 \int\!\!\int 
r_{12}^{-1}
\rho_1(\mathbf{r}_1,\mathbf{r}_2) 
\rho_1(\mathbf{r}_2,\mathbf{r}_1) 
\,d\mathbf{r}_1
\,d\mathbf{r}_2,
\end{eqnarray}
and the particle density is
\begin{equation}\label{den} %\mbox{\tiny den}
\rho(\mathbf{r})  = \rho_1(\mathbf{r},\mathbf{r}). 
\end{equation}

\subsection{Hartree--Fock Theory}

Using Eqs.~(\ref{kernexch}) and (\ref{gamma.sf}) it is readily observed that the
kernel of the exchange operator is given by
\begin{eqnarray} \label{kernexchpot.sf} %\mbox{\tiny kernexchpot.sf}
v_{\mathrm{x}}^{\scscs \gamma}(\mathbf{x}_1,\mathbf{x}_2)=
v_{\mathrm{x}}^{\scscs \rho_1}(\mathbf{r}_1,\mathbf{r}_2)\delta_{\omega_1,\omega_2},
\end{eqnarray}
where
\begin{eqnarray} \label{exchpot.kern} %\mbox{\tiny exchpot.kern}
v_{\mathrm{x}}^{\scscs \rho_1}(\mathbf{r}_1,\mathbf{r}_2)= 
\mbox{ \large $-\frac12$} r_{12}^{-1}
\rho_1(\mathbf{r}_1,\mathbf{r}_2).
\end{eqnarray}
Therefore, for an arbitrary function, say $\phi$, we have
\begin{eqnarray}
\hat{v}_{\mathrm{x}}^{\scscs \gamma} \phi(\mathbf{x}_1)=
\hat{v}_{\mathrm{x}}^{\scscs \rho_1} \phi(\mathbf{x}_1),
\end{eqnarray}
where $v_{\mathrm{x}}^{\scscs \gamma}(\mathbf{x}_1,\mathbf{x}_2)$ and $v_{\mathrm{x}}^{\scscs
\rho_1}(\mathbf{r}_1,\mathbf{r}_2)$ serve as the kernels of $\hat{v}_{\mathrm{x}}^{\scscs
\gamma}$ and $\hat{v}_{\mathrm{x}}^{\scscs \rho_1}$, respectively:
\begin{eqnarray}
\hat{v}_{\mathrm{x}}^{\scscs \gamma} \phi(\mathbf{x}_1)&=& 
\int v_{\mathrm{x}}^{\scscs \gamma}(\mathbf{x}_1,\mathbf{x}_2) \phi(\mathbf{x}_2) 
\; d \mathbf{x}_2,\\ 
\label{exchpot.ident} %\mbox{\tiny exchpot.ident}
\hat{v}_{\mathrm{x}}^{\scscs \rho_1} \phi^\prime(\mathbf{r}_1) &=&
\int v_{\mathrm{x}}^{\scscs \rho_1}(\mathbf{r}_1,\mathbf{r}_2)
\phi^\prime(\mathbf{r}_2) \; d \mathbf{r}_2.
\end{eqnarray}

Using Eqs.~(\ref{kernexchpot.sf}) and (\ref{gamma.sf}), the kernel of the Fock
operators, Eq.\ (\ref{kernfock}), can be written as
\begin{subequations}
\begin{equation} \label{fock.sf} %\mbox{\tiny fock.sf}
F_\gamma(\mathbf{x}_1,\mathbf{x}_2) = 
\delta_{\omega_1\omega_2}  F_{\rho_1}(\mathbf{r}_1,\mathbf{r}_2),
\end{equation}
where
\begin{equation} \label{kernfock.sf} %\mbox{\tiny kernfock.sf}
F_{\rho_1}(\mathbf{r}_1,\mathbf{r}_2)
=
\delta(\mathbf{r}_2-\mathbf{r}_1) 
\left(
-\mbox{\small$\frac{1}{2}$}
\nabla_{\!\mbox{\tiny $2$}}^2
+ v(\mathbf{r}_2)
+
\int 
r_{23}^{-1}
\rho(\mathbf{r}_3) 
\,d\mathbf{r}_3\right)
+
v_{\mathrm{x}}^{\scscs \rho_1}(\mathbf{r}_1,\mathbf{r}_2),
\end{equation}
\end{subequations}
Substituting Eqs.~(\ref{spinorb.res}) and (\ref{fock.sf}) into (\ref{F.can}), and
summing over $\omega_2$ we get
\begin{eqnarray} \label{F.can.sf} %\mbox{\tiny F.can.sf}
\int F_{\tilde{\varrho}_1}(\mathbf{r}_1,\mathbf{r}_2)
\chi_j^{\tilde{\varrho}_1}(\mathbf{r}_2)
\; d \mathbf{r}_2 \sigma(\omega_1)
&=&
\varepsilon_{j}^{\mbox{\tiny $\tilde{\varrho}_1$}}
\chi^{\tilde{\varrho}_1}_j(\mathbf{r}_1)\sigma(\omega_1).
\end{eqnarray}
By letting $F_{\tilde{\varrho}_1}(\mathbf{r}_1,\mathbf{r}_2)$ serve as the kernel of the
spin-free Fock-operator $\hat{F}_{\tilde{\varrho}_1}$, we have
\begin{eqnarray} \label{F.canc.sf} %\mbox{\tiny F.canc.sf}
\hat{F}_{\tilde{\varrho}_1}
\chi_j^{\tilde{\varrho}_1}(\mathbf{r}_1)
&=&
\varepsilon_{j}^{\mbox{\tiny $\tilde{\varrho}_1$}}
\chi^{\tilde{\varrho}_1}_j(\mathbf{r}_1),
\end{eqnarray}
where $\tilde{\varrho}_1$ is the spinless one-particle density-matrix of the
Hartree-Fock state:
\begin{equation} \label{hftau.sf} %\mbox{\tiny hftau.sf}
\tilde{\tau}(\mathbf{x}_1,\mathbf{x}_2) = \frac12 \tilde{\varrho}_1(\mathbf{r}_1,\mathbf{r}_2)
\delta_{\omega_1\omega_2},
\end{equation}
and the Hartree--Fock spatial-orbitals are denoted in an analogous way as the
spin-orbitals from Sec.~(\ref{HFT}), e.g., $\tilde{\varrho_1}$ replaces
$\tilde{\tau}$.

Substituting Eqs.~(\ref{kernfock.sf}) into (\ref{F.can.sf}) and using
(\ref{exchpot.ident}) gives
\begin{eqnarray} \label{F.canb.sf} %\mbox{\tiny F.canb.sf}
\left(
-\mbox{\small$\frac{1}{2}$}
\nabla_{\mbox{\tiny $1$}}^2
+ v(\mathbf{r}_1)
+
\int 
r_{12}^{-1}
\tilde{\varrho}(\mathbf{r}_2) 
\,d\mathbf{r}_2
- \hat{v}_{\mathrm{x}}^{\scscs \tilde{\varrho}_1}(\mathbf{r}_1)
\right) 
\chi^{\tilde{\varrho}_1}_i(\mathbf{r}_1)
=
\varepsilon_{i}^{\mbox{\tiny $\tilde{\varrho}_1$}}
\chi^{\tilde{\varrho}_1}_i(\mathbf{r}_1).
\end{eqnarray}
Eqs.~(\ref{F.canc.sf}) and (\ref{F.canb.sf}) are the closed-shell spin-free forms of
Eqs.~(\ref{F.can}) and (\ref{F.canb}); the spin-free forms or Eqs.~(\ref{kernfock})
and (\ref{kernexch}) are easily proven to satisfy the following relations:
\begin{eqnarray} \label{kernfock.def.sf} %\mbox{\tiny kernfock.def.sf}
F_{\rho_1}(\mathbf{r}_1,\mathbf{r}_2)&=&
\frac{\delta E_1[\rho_1]} 
{\delta \rho_1(\mathbf{r}_2,\mathbf{r}_1)}, \\
\label{kernexch.sf} %\mbox{\tiny kernexch.sf}
v_{\mathrm{x}}^{\scscs \rho_1}(\mathbf{r}_1,\mathbf{r}_2)
&=&
\frac{\delta E_{\mathrm{x}}[\rho_1]} 
{\delta \rho_1(\mathbf{r}_2,\mathbf{r}_1)}, 
\end{eqnarray}
where, as mentioned previously, these functions are given by Eqs,~(\ref{kernfock.sf})
and (\ref{exchpot.kern}), respectively.

\subsection{Brueckner orbital theory} 

We now generalize many of the Hartree--Fock relations from the previous subsection
simply by replacing the exchange operator $\hat{v}_{\mathrm{x}}^{\scscs \gamma}$
with the exchange-correlation operator $\hat{\nu}_{\mathrm{xc}}^{\scscs \gamma\eta}$.

For the closed shell case under consideration, it can be easily demonstrated that the
kernels of the exchange-correlation operators are given by \cite{tobe}
\begin{eqnarray} \label{kernexchcorrpot.sf} %\mbox{\tiny kernexchcorrpot.sf}
\nu_{\mathrm{xc}}^{\scscs \gamma\eta}(\mathbf{x}_1,\mathbf{x}_2)=
\nu_{\mathrm{xc}}^{\scscs \rho_1\!\eta}(\mathbf{r}_1,\mathbf{r}_2)
\delta_{\omega_1,\omega_2}.
\end{eqnarray}
Therefore, for an arbitrary function, say $\phi$, we have
\begin{eqnarray}
\hat{\nu}_{\mathrm{xc}}^{\scscs \gamma\eta} \phi(\mathbf{x}_1)=
\hat{\nu}_{\mathrm{xc}}^{\scscs \rho_1\!\eta} \phi(\mathbf{x}_1),
\end{eqnarray}
where $\nu_{\mathrm{xc}}^{\scscs \gamma\eta}(\mathbf{x}_1,\mathbf{x}_2)$ and
$\nu_{\mathrm{xc}}^{\scscs \rho_1\!\eta}(\mathbf{r}_1,\mathbf{r}_2)$ serve as the
kernels of $\hat{\nu}_{\mathrm{xc}}^{\scscs \gamma\eta}$ and
$\hat{\nu}_{\mathrm{xc}}^{\scscs \rho_1\!\eta}$, respectively.

Using Eqs.~(\ref{kernexchcorrpot.sf}) and (\ref{gamma.sf}), the kernel of the
exact Fock operators, Eq.\ (\ref{kernEfock}), can be written as
\begin{subequations}
\begin{equation} \label{efock.sf} %\mbox{\tiny efock.sf}
{\cal \zeta}_{\gamma}^{\scscs (\eta)}(\mathbf{x}_1,\mathbf{x}_2) =
\delta_{\omega_1\omega_2}  
{\cal \zeta}_{\rho_1}^{\scscs (\eta)}(\mathbf{r}_1,\mathbf{r}_2),
\end{equation}
where
\begin{equation} \label{kernEfock.sf} %\mbox{\tiny kernEfock.sf}
{\cal \zeta}_{\rho_1}^{\scscs (\eta)}(\mathbf{r}_1,\mathbf{r}_2)
=
\delta(\mathbf{r}_2-\mathbf{r}_1) 
\left(
-\mbox{\small$\frac{1}{2}$}
\nabla_{\!\mbox{\tiny $2$}}^2
+ v(\mathbf{r}_2)
+
\int 
r_{23}^{-1}
\rho(\mathbf{r}_3) 
\,d\mathbf{r}_3\right)
+
\nu_{\mathrm{xc}}^{\scscs \rho_1\!\eta}(\mathbf{r}_1,\mathbf{r}_2).
\end{equation}
\end{subequations}
Substituting Eqs.~(\ref{spinorb.res}) and (\ref{efock.sf}) into (\ref{eF.can}),
and summing over $\omega_2$ we get
\begin{eqnarray} \label{eF.can.sf} %\mbox{\tiny eF.can.sf}
\int 
{\cal \zeta}_{\varrho_1}^{\scscs (\eta)}
(\mathbf{r}_1,\mathbf{r}_2)
\chi_j^{\varrho_1}(\mathbf{r}_2)
\; d \mathbf{r}_2 \sigma(\omega_1)
&=&
\xi_{j}^{\mbox{\tiny $\varrho_1$}}
\chi^{\varrho_1}_j(\mathbf{r}_1)\sigma(\omega_1).
\end{eqnarray}
By letting ${\cal \zeta}_{\varrho_1}^{\scscs (\eta)}(\mathbf{r}_1,\mathbf{r}_2)$
serve as the kernels of the exact, spin-free Fock-operator ${\cal
\hat{\zeta}}_{\varrho_1}^{\scscs (\eta)}$, we have
\begin{eqnarray} \label{eF.canc.sf} %\mbox{\tiny eF.canc.sf}
{\cal \hat{\zeta}}_{\varrho_1}^{\scscs (\eta)}
\chi_j^{\varrho_1}(\mathbf{r}_1)
&=&
\xi_{j}^{\mbox{\tiny $\varrho_1$}}
\chi^{\varrho_1}_j(\mathbf{r}_1),
\end{eqnarray}
where $\varrho_1$ is the spinless one-particle density of the Brueckner state:
\begin{equation} \label{tau.sf} %\mbox{\tiny tau.sf}
\tau(\mathbf{x}_1,\mathbf{x}_2) = \frac12 \varrho_1(\mathbf{r}_1,\mathbf{r}_2)
\delta_{\omega_1\omega_2},
\end{equation}
and the Brueckner spatial-orbitals are denoted in an analogous way as the
spin-orbitals from Sec.~(\ref{VBOT}), e.g., $\varrho_1$ replaces $\tau$.

Substituting Eqs.~(\ref{kernEfock.sf}) into (\ref{eF.can.sf}) gives 
\begin{eqnarray} \label{eF.canb.sf} %\mbox{\tiny eF.canb.sf}
\left(
-\mbox{\small$\frac{1}{2}$}
\nabla_{\mbox{\tiny $1$}}^2
+ v(\mathbf{r}_1)
+
\int 
r_{12}^{-1}
\varrho(\mathbf{r}_2) 
\,d\mathbf{r}_2
+ \hat{\nu}_{\mathrm{xc}}^{\scscs \varrho_1\!\eta}(\mathbf{r}_1)
\right) 
\chi^{\varrho_1}_i(\mathbf{r}_1)
=
\xi_{i}^{\mbox{\tiny $\varrho_1$}}
\chi^{\varrho_1}_i(\mathbf{r}_1).
\end{eqnarray}
Eqs.~(\ref{eF.canc.sf}) and (\ref{eF.canb.sf}) are the closed-shell spin-free
forms of Eqs.~(\ref{eF.can}) and (\ref{eF.canb}); the spin-free forms or
Eqs.~(\ref{kernEfock}) and (\ref{kernexchcorr}) can be shown to satisfy the
following relations \cite{tobe}:
\begin{eqnarray} \label{kernfock.def.sfB} %\mbox{\tiny kernfock.def.sfB}
{\cal \zeta}_{\rho_1}^{\scscs (\eta)}(\mathbf{r}_1,\mathbf{r}_2)&=&
\frac{\delta \bar{E}_\eta[\rho_1]} 
{\delta \rho_1(\mathbf{r}_2,\mathbf{r}_1)}, \\
\label{kernexchcorr.sf} %\mbox{\tiny kernexchcorr.sf}
\nu_{\mathrm{xc}}^{\scscs \rho_1\!\eta}(\mathbf{r}_1,\mathbf{r}_2)
&=&
\frac{\delta \bar{E}^{\scscs (\eta)}_{\mathrm{xc}}[\rho_1]} 
{\delta \rho_1(\mathbf{r}_2,\mathbf{r}_1)}.
\end{eqnarray}

Substituting Eq.~(\ref{exco.sfv}) into (\ref{kernexchcorr.sf}) and using 
(\ref{kernexch.sf}), we have
\begin{eqnarray} \label{kernels} %\mbox{\tiny kernels}
\nu_{\mathrm{xc}}^{\scscs \rho_1\!\eta}(\mathbf{r}_1,\mathbf{r}_2)
=
v_{\mathrm{x}}^{\scscs \rho_1}(\mathbf{r}_1,\mathbf{r}_2)
+
\nu_{\mathrm{co}}^{\scscs \rho_1\!\eta}(\mathbf{r}_1,\mathbf{r}_2),
\end{eqnarray}
where the kernel of the exchange potential $v_{\mathrm{x}}^{\scscs
\rho_1\!\eta}(\mathbf{r}_1,\mathbf{r}_2)$ is given by Eq.~(\ref{exchpot.kern});
the kernel of the correlation potential $\nu_{\mathrm{co}}^{\scscs
\rho_1\!\eta}(\mathbf{r}_1,\mathbf{r}_2)$ is defined by
\begin{eqnarray} \label{Eco.funder} %\mbox{\tiny Eco.funder}
\nu_{\mathrm{co}}^{\scscs \rho_1\!\eta}(\mathbf{r}_1,\mathbf{r}_2)
=
\frac{\delta \bar{E}^{\scscs (\eta)}_{\mathrm{co}}[\rho_1]} 
{\delta \rho_1(\mathbf{r}_2,\mathbf{r}_1)}.
\end{eqnarray}
The operator form of Eq.~(\ref{kernels}) is given by
\begin{eqnarray} \label{operators} %\mbox{\tiny operators}
\hat{\nu}_{\mathrm{xc}}^{\scscs \rho_1\!\eta}
=
\hat{v}_{\mathrm{x}}^{\scscs \rho_1}
+
\hat{\nu}_{\mathrm{co}}^{\scscs \rho_1\!\eta}.
\end{eqnarray}

\section{Approximations} \label{approx}

The target state $|\Psi\rangle$ can be generated from a reference state
$|\Phi\rangle$ by its wave-operator~$\Omega_\Phi$:
\begin{eqnarray}
|\Psi\rangle = \Omega_\Phi|\Phi\rangle = |\Psi_\Phi\rangle + 
|\Psi_{\Phi,\Theta}\rangle, 
\end{eqnarray}
where we have partitioned the target state into two terms: $|\Psi_\Phi\rangle$ and
$|\Psi_{\Phi,\Theta}\rangle$. Consider the diagrammatic perturbation-expansion for
$\Omega_\Phi|\Phi\rangle$, $|\Psi_{\Phi,\Theta}\rangle$ is then defined as the sum
of {\em all} diagrammatic terms from the perturbation expansion that must vanish
for any external potential when the reference state is the Brueckner one:
\begin{eqnarray}
|\Psi_{\Theta,\Theta}\rangle = 0.
\end{eqnarray}
Hence, $|\Psi_\Phi\rangle$ is a trial wavefunction that fulfills all of the
requirements of the ones considered in Sec.~\ref{VBOT}; furthermore, by our
definition, $|\Psi_\Phi\rangle$ is the best trial wavefunction, since it contains
no extra terms that must vanish when ($|\Phi\rangle=|\Theta\rangle$).

Consider a system with an external potential, say $v^\prime$, where its
correlation energy ${\cal E}_{\mathrm{co}}$ is known when the reference state is
the Brueckner one with a corresponding one-particle density-matrix of
$\tau^\prime$; using Eq.~(\ref{co=cof}) we have
\begin{eqnarray} %\label{co=cof} \mbox{\tiny co=cof}
{\cal E}_{\mathrm{co}}[\tau^\prime,v^\prime]&=&
\bar{E}_{\mathrm{co}}[\tau^\prime,v^\prime],
\end{eqnarray}
where the correlation-energy functional from $|\Psi_\Phi\rangle$ is denoted by
$\bar{E}_{\mathrm{co}}$ and we have indicated that these functionals depend on the
external potential.  Hence, a reasonable approximation for $\bar{E}_{\mathrm{co}}$
at other potentials $v$ and one-particle density-matrices $\gamma$ is given by
\begin{eqnarray} \label{approx5} %\mbox{\tiny approx5}
\bar{E}_{\mathrm{co}}[\gamma,v]&=&
{\cal E}_{\mathrm{co}}[\tau^\prime,v^\prime]_{(\tau^\prime=\gamma,v^\prime=v)}.
\end{eqnarray}

Now consider the four trial wavefunctions which can also be obtained using a wave
operator:
\begin{eqnarray}
|\Psi_\Phi^{\scscs(\eta)}\rangle = \Omega_\Phi^{\scscs(\eta)}|\Phi\rangle,
\end{eqnarray}
where from the previous notation we have ($\Omega_\Phi^{\scscs(\mathrm{III})} =
\tilde{\Omega}_\Phi$). Consider the first two cases: ($\eta = \mbox{{\small I} and
{\small II}}$), where the wave-operators expansion are easily obtained by
considering Lindgren's formalism \cite{Lindgren:74,Lindgren:78,Lindgren:86}.  By
examining this expansion, it is readily seen that the trial wavefunctions 
contain terms, say $|\Psi_{\Phi,\Theta}^{^{\scscs(\eta)}}\rangle$, that, by
necessity, vanish when ($|\Phi\rangle=|\Theta\rangle$):
\begin{eqnarray}
|\Psi_\Phi^{\scscs(\eta)}\rangle = |\Psi_\Phi^{\scscs(\eta)}\rangle + 
|\Psi^{\scscs(\eta)}_{\Phi,\Theta}\rangle.
\end{eqnarray}
Hence, even though the approximation (\ref{approx5}) is reasonable for ($\eta =
\mbox{{\small I} and {\small II}}$):
\begin{eqnarray} \label{approx6} %\mbox{\tiny approx6}
\bar{E}_{\mathrm{co}}^{\scscs(\eta)}[\gamma,v]&=&
{\cal E}_{\mathrm{co}}[\tau^\prime,v^\prime]_{(\tau^\prime=\gamma,v^\prime=v)},
\end{eqnarray}
it neglects the contribution to $\bar{E}_{\mathrm{co}}^{\scscs(\eta)}$ arising
from $|\Psi^{\scscs(\eta)}_{\Phi,\Theta}\rangle$.

The wave operator diagrams for ($\eta = \mbox{\small III}$) are more complicated
to describe and are given elsewhere \cite{tobe}. However, it is obvious from the
previous work \cite{Finley:bdmt} that the term
$|\Psi^{\scscs(\mathrm{III})}_{\Phi,\Theta}\rangle$ is considerably smaller in
magnitude than for ($\eta = \mbox{{\small I} and {\small II}}$), suggesting that
the approximation (\ref{approx6}) is reasonable, where this type of approximation
was considered previously \cite{Finley:bdmt}. It is quite possible that
$|\Psi^{\scscs(\mathrm{III})}_{\Phi,\Theta}\rangle$ is zero, so, in that case
($|\Psi^{\scscs(\mathrm{III})}_{\Phi}\rangle = |\Psi_{\Phi}\rangle$), this was
previously presumed to be true \cite{Finley:bdmt}, but it was not discussed in
detail.

The wave-operator diagrams for ($\eta = \mbox{\small IV}$) contain unlinked terms
and will not be considered, except to mention that
$|\Psi^{\scscs(\mathrm{IV})}_{\Phi,\Theta}\rangle$ is nonzero.

\subsection{Local Density Approximation} \label{lda}

The correlation energy of an electron gas -- denoted by ${\cal
E}_{\mathrm{co}}^{\scscs (\text{gas})}$ -- is given by the correlation energy
${\cal E}_{\mathrm{co}}$ for a system where the external potential $v$ is
constant. Since the orbitals for an electron gas are Brueckner, and using
Eq.~(\ref{co=cof}), we have
\begin{eqnarray}\label{Egas.corrE} %\mbox{\tiny Egas.corrE}
{\cal E}_{\mathrm{co}}^{\scscs (\text{gas})} [\tau_g] =
{\cal E}_{\mathrm{co}}[\tau_g,v]
= \bar{E}_{\mathrm{co}}^{\scscs (\eta)}[\tau_g,v], \; \; v=\mbox{const},
\end{eqnarray}
where $\tau_g$ is the one-particle density-matrix that is from the Brueckner
reference-state of an electron gas, say $|\Theta_g\rangle$, where the Brueckner
orbitals, defining $|\Theta_g\rangle$, are given by plane waves when periodic
boundary conditions are imposed; furthermore, the correlation energy ${\cal
E}_{\mathrm{co}}$ is unchanged if $v$ is changed by a constant amount; ${\cal
E}_{\mathrm{co}}^{\scscs (\text{gas})} [\tau_g]$ does not depend on $v$. (A
constant shift in $v$ does, however, shift the first-order energy $E_1$.)

The electron-gas approximation for the third correlation-energy functional
utilizes this functional, and it is given by \cite{Finley:bdmt}:
\begin{eqnarray}\label{Egas.approxB.v} %\mbox{\tiny Egas.approxB.v}
\bar{E}_{\mathrm{co}}^{\scscs (\mathrm{III})}[\gamma]&\approx& {\cal
E}_{\mathrm{co}}^{\scscs (\text{gas})} [\tau_g]_{(\tau_{\mbox{\tiny$g$}}=\gamma)},
\end{eqnarray}
and this approximation is a special case of Eq.~(\ref{approx6}), where there are
no terms involving an external potential. (We can also use this approximation for
$\bar{E}_{\mathrm{co}}$ defined in the previous subsection.)

For spatially restricted orbitals, given by Eq.~(\ref{spinorb.res}), the above
Eq.\ becomes
\begin{eqnarray}\label{Egas.approxB.v2} %\mbox{\tiny Egas.approxB.v2}
\bar{E}_{\mathrm{co}}^{\scscs (\mathrm{III})}[\rho_1]&\approx& {\cal
E}_{\mathrm{co}}^{\scscs (\text{gas})} [\varrho_{1g}]_{(\varrho_{\mbox{\tiny$1g$}}=\rho_1)},
\end{eqnarray}
where $\varrho_{1g}$ is the spin-less, one-particle density-matrix of the
Brueckner reference-state for an electron gas:
\begin{equation} \label{onebr.sf} %\mbox{\tiny onebr.sf}
\tau_g(\mathbf{x}_1,\mathbf{x}_2) = \frac12 \varrho_{1g}(\mathbf{r}_1,\mathbf{r}_2)
\delta_{\omega_1\omega_2}.
\end{equation}

For a {\em uniform} electron gas, Eq.~(\ref{onebr.sf}) is denoted by
\begin{equation} %\label{} \mbox{\tiny }
\tau_{\text{ug}}(\mathbf{x}_1,\mathbf{x}_2) = \frac12
\varrho_{\text{1ug}}(\mathbf{r}_1,\mathbf{r}_2) \delta_{\omega_1\omega_2},
\end{equation}
and the density that is from the Brueckner, one-particle reference-state of a
uniform electron gas, say $|\Theta_{\text{ug}}\rangle$, is given by
\begin{equation} \label{refden.ug} %\mbox{\tiny refden.ug}
\varrho_{\text{ug}}(\mathbf{r}) =
\varrho_{\text{1ug}}(\mathbf{r},\mathbf{r}),
\end{equation}
and this density is constant, i.e., it does not depend on $\mathrm{r}$;
furthermore, we have
\begin{equation} \label{den.ug} %\mbox{\tiny den.ug}
\varrho_{\text{ug}} = n_{\text{ug}},
\end{equation}
where $n_{\text{ug}}$ is the density from the target state, say
$|\Psi_{\text{ug}}\rangle$, of a uniform electron gas:
\begin{equation} \label{trwf=exact.ug} %\mbox{\tiny trwf=exact.ug}
|\Psi_{\Theta_{\text{ug}}}^{\scscs (\eta)}\rangle = |\Psi_{\text{ug}}\rangle,
\end{equation}
where we have used Eq.~(\ref{trwf=exact}).

The correlation energy ${\cal E}_{\mathrm{co}}^{\scscs (\text{ug})}$ of a uniform
electron gas depends on the constant density $n_{\text{ug}}$; it is given by the
following relation:
\begin{eqnarray} \label{Ecorr.ug} %\mbox{\tiny Ecorr.ug}
{\cal E}_{\mathrm{co}}^{\scscs (\text{ug})} (n_{\text{ug}})=
N \epsilon_{\mathrm{co}}^{\scscs \text{u\hspace{-0.3ex}g}} (n_{\text{ug}}),
\end{eqnarray}
where $\epsilon_{\mathrm{co}}^{\scscs \text{u\hspace{-0.3ex}g}}$ is the
correlation-energy per particle and $N$ is the number of particles. Using
Eq.~(\ref{den.ug}) gives
\begin{eqnarray} \label{Ecorr.ugB} %\mbox{\tiny Ecorr.ugB}
{\cal E}_{\mathrm{co}}^{\scscs (\text{ug})} (\varrho_{\text{ug}})=
N \epsilon_{\mathrm{co}}^{\scscs \text{u\hspace{-0.3ex}g}} (\varrho_{\text{ug}}).
\end{eqnarray}
Following Kohn and Sham \cite{Kohn:65}, but using Eq.~(\ref{Ecorr.ugB}) instead of
(\ref{Ecorr.ug}), we construct the following functional for an electron gas:
\begin{equation} \label{ldaC} %\mbox{\tiny ldaC}
{\cal E}_{\mathrm{co}}^{\scscs (\text{gas})} [\varrho_g] \approx \int \varrho_g(\mathbf{r})
\hspace{0.2ex} \epsilon_{\mathrm{co}}^{\scscs \text{u\hspace{-0.3ex}g}} (\varrho_g) \;
d \mathbf{r}.
\end{equation}
This functional can be expressed as one that appears to depend upon the
one-particle density-matrix:
\begin{equation} \label{ldaD} %\mbox{\tiny ldaD}
{\cal E}_{\mathrm{co}}^{\scscs (\text{gas})} [\varrho_{1g}] \approx \int\!\int
\varrho_{1g}(\mathbf{r}_2,\mathbf{r}_1)\delta(\mathbf{r}_1-\mathbf{r}_2)
\epsilon_{\mathrm{co}}^{\scscs \text{u\hspace{-0.3ex}g}} 
(\varrho_{1g}(\mathbf{r}_2,\mathbf{r}_1)) 
\; d \mathbf{r}_1
\; d \mathbf{r}_2,
\end{equation}
where ($\varrho_{1g}(\mathbf{r},\mathbf{r})= \varrho_{g}(\mathbf{r})$).
Substituting the above expression into Eq.~(\ref{Egas.approxB.v2}) gives
\begin{equation} \label{ldaE} %\mbox{\tiny ldaE}
\bar{E}_{\mathrm{co}}^{\scscs (\mathrm{III})}[\rho_1]
\approx \int \int
\rho_1(\mathbf{r}_2,\mathbf{r}_1)\delta(\mathbf{r}_1-\mathbf{r}_2)
\epsilon_{\mathrm{co}}^{\scscs \text{u\hspace{-0.3ex}g}} 
(\rho_1(\mathbf{r}_2,\mathbf{r}_1)) 
\; d \mathbf{r}_1
\; d \mathbf{r}_2.
\end{equation}

The correlation potential $\hat{\nu}_{\mathrm{co}}^{\scscs \rho_1\!\eta}$ is
defined by its kernel, given by Eq.~(\ref{Eco.funder}). Because of the
$\delta(\mathbf{r}_1-\mathbf{r}_2)$ factor, $\hat{\nu}_{\mathrm{co}}^{\scscs
\rho_1\!\mbox{\tiny $\mathrm{III}$}}$ is a local operator when using the above
expressions for $\bar{E}_{\mathrm{co}}^{\scscs (\mathrm{III})}$; explicitly, using
the results from Appendix \ref{pot.ident}, we have the following approximate
expressions:
\begin{eqnarray} \label{ldaF} %\mbox{\tiny ldaF}
\bar{E}_{\mathrm{co}}^{\scscs (\mathrm{III})}[\rho_1]&\approx&
E_{\mathrm{co}}^{\mathrm{ug}}[\rho]
=\int 
\rho(\mathbf{r})
\epsilon_{\mathrm{co}}^{\scscs \text{u\hspace{-0.3ex}g}} (\rho) \;
d \mathbf{r},\\
%\label{} \mbox{\tiny }
\hat{\nu}_{\mathrm{co}}^{\scscs \rho_1\!\mbox{\tiny $\mathrm{III}$}}(\mathbf{r}) &\approx&
\nu_{\mathrm{co}}^{\scscs \mathrm{ug}}(\mathbf{r})= 
\frac{\delta E_{\mathrm{co}}^{\mathrm{ug}}[\rho]} 
{\delta \rho(\mathbf{r})},
\end{eqnarray}
where, for future reference, these approximations are denoted by
$E_{\mathrm{co}}^{\mathrm{ug}}$ and $\nu_{\mathrm{co}}^{\scscs \mathrm{ug}}$.

Using the exchange-energy formula by Dirac for a uniform electron gas, the
exchange portion can be treated in an identical manner as the correlation; this
approach yields the following local exchange-operator and exchange energy:
\begin{eqnarray} %\label{} \mbox{\tiny }
E_{\mathrm{x}}[\rho_1]&\approx& E_{\mathrm{x}}^{\mathrm{ug}}[\rho]
= \frac{3}{4}\left(\frac{3}{\pi}\right)^{1/3} \int 
\rho(\mathbf{r})^{4/3}\; d \mathbf{r},\\
\label{local.exch} %\mbox{\tiny local.exch}
\hat{v}_{\mathrm{x}}^{\scscs \rho_1}(\mathbf{r}) 
&\approx&  v_{\mathrm{x}}^{\scscs \mathrm{ug}}(\mathbf{r})  =
\frac{\delta E_{\mathrm{x}}^{\mathrm{ug}}[\rho]} 
{\delta \rho(\mathbf{r})}=
\left(\frac{3}{\pi}\rho(\mathbf{r})\right)^{1/3},
\end{eqnarray} 
where $E_{\mathrm{x}}^{\mathrm{ug}}$ denotes the Dirac exchange-energy functional
and $v_{\mathrm{x}}^{\scscs \mathrm{ug}}$ is the corresponding exchange potential.
The above four expressions are identical to the ones for the LDA of DFT, except
that -- in the current approach -- $\rho$ is {\em not} required to be the density
of an exact wavefunction $|\Psi\rangle$; $\rho$ is {\em only} required to be the
density of a reference state $|\Phi\rangle$, and this is assured to be the case.

\subsection{The Becke exchange correction}

For closed-shell systems that are well described by a single determinant, it is
well known that the Bruckner and Hartree--Fock reference-states, $|\Theta\rangle$
and $|\tilde{\Theta}\rangle$, are usually very similar. Hence, in these cases, the
spinless, one-particle density-matrices from the Brueckner reference-state,
$\varrho_1$, is similar to the one from the Hartree--Fock reference state,
$\tilde{\varrho}_1$. Therefore, an approximate exchange-functional
$E_{\mathrm{x}}[\rho_1]$ that depends on the density of the reference state, i.e.,
\begin{eqnarray} \label{exch.approx} %\mbox{\tiny exch.approx}
E_{\mathrm{x}}[\rho_1]\approx
\tilde{E}_{\mathrm{x}}[\rho], 
\end{eqnarray}
and is an accurate approximation with the Hartree--Fock approach, is also a
reasonable approximation for the current approach.  Of particular interest is the
exchange functional obtained by Becke \cite{Becke:88} -- where this functional is
an approximation of the Hartree--Fock exchange with empirical parameters fit to
atomic systems; explicitly, this functional is given by
\begin{eqnarray} \label{becke88} %\mbox{\tiny becke88}
\tilde{E}_{\mathrm{x}}[\rho] = E_{\mathrm{x}}^{\mathrm{ug}}[\rho] + \Delta
\!E_{\mathrm{x}}^{\scscs \mathrm{B88}}[\rho],
\end{eqnarray}
where $\Delta \!E_{\mathrm{x}}^{\scscs \mathrm{B88}}$ is the Becke exchange
correction to the uniform electron-gas term, $E_{\mathrm{x}}^{\mathrm{ug}}$.
Since this correction term contains a gradient dependence, the general form of the
functional can be written using the following integral relation:
\begin{eqnarray} \label{becke88B} %\mbox{\tiny becke88B}
\tilde{E}_{\mathrm{x}}[\rho] =
\int G_{\mathrm{x}}(\rho,\nabla\rho)\, d\mathbf{r}.
\end{eqnarray} 
By using Eqs.~(\ref{exch.approx}) and (\ref{becke88}) and the
following identity:
\begin{eqnarray} \label{ident.grad} %\mbox{\tiny ident.grad}
\nabla_1\rho(\mathbf{r}_1) = 
\left[(\nabla_1 + \nabla_2)\rho_1(\mathbf{r}_2,\mathbf{r}_1
)\right]_{\mathbf{r}_2=\mathbf{r}_1},
\end{eqnarray}
we obtain the following approximation: 
\begin{eqnarray} \label{b88} %\mbox{\tiny b88}
E_{\mathrm{x}}[\rho_1] \approx E_{\mathrm{x}}^{\mathrm{ug}}[\rho] + \Delta
\!E_{\mathrm{x}}^{\scscs \mathrm{B88}}[\rho] =
\int \!\!\int \delta(\mathbf{r}_2-\mathbf{r}_1)
G_{\mathrm{x}}
\left(\makebox(0,11){}
\rho_1(\mathbf{r}_2,\mathbf{r}_1),
(\nabla_1 + \nabla_2)\rho_1(\mathbf{r}_2,\mathbf{r}_1)
\right)
\, d\mathbf{r}_1 \, d\mathbf{r}_2. \nonumber 
%\\
\end{eqnarray}
Using this expression for the exchange energy, the exchange potential is local;
using the result from Appendix \ref{pot.ident} and the definition for
$v_{\mathrm{x}}^{\scscs \mathrm{ug}}$ given by Eq.~(\ref{local.exch}), we have
\begin{eqnarray} \label{B88.pot} %\mbox{\tiny B88.pot}
\hat{v}_{\mathrm{x}}^{\scscs \rho_1}(\mathbf{r}) 
&\approx&  v_{\mathrm{x}}^{\scscs \mathrm{ug}}(\mathbf{r}) +
v_{\mathrm{x}}^{\scscs \mathrm{B88}}(\mathbf{r})  =
\frac{\delta E_{\mathrm{x}}^{\mathrm{ug}}[\rho]} 
{\delta \rho(\mathbf{r})}
+
\frac{\delta \Delta \!E_{\mathrm{x}}^{\scscs \mathrm{B88}}[\rho]} 
{\delta \rho(\mathbf{r})},
\end{eqnarray}
where this expression defines $v_{\mathrm{x}}^{\scscs \mathrm{B88}}(\mathbf{r})$.

\subsection{The LYP Functional}

Consider the approximation given by Eq.~(\ref{approx6}), where ($\eta=
\mbox{{\small III}}$), and the correlation energy ${\cal E}_{\mathrm{co}}$ -- which
is also a functional -- is taken to be the one from the helium atom:
\begin{eqnarray} \label{approx7} %\mbox{\tiny approx7}
\bar{E}_{\mathrm{co}}^{\scscs(\mathrm{III})}[\gamma,v]&=&
{\cal E}_{\mathrm{co}} 
[\tau_{\mbox{\tiny\textsc{h}e}},v_{\mbox{\tiny\textsc{h}e}}]
_{(\tau_{\mbox{\tiny\textsc{h}e}} = \gamma,\,v_{\mbox{\tiny\textsc{h}e}} = v)},
\end{eqnarray}
where $\tau_{\mbox{\tiny\textsc{h}e}}$ is the Brueckner, one-particle,
density-matrix for the helium atom, and $v_{\mbox{\tiny\textsc{h}e}}$ is the
external potential for this system.  Furthermore, since the Brueckner
density-matrix $\tau_{\mbox{\tiny\textsc{h}e}}$ is approximately equal to the
Hartree--Fock one, say $\tilde{\tau}_{\mbox{\tiny\textsc{h}e}}$, we can write
\begin{equation} \label{He.approxC} %\mbox{\tiny He.approxC}
E_{\mathrm{co}}^{\scscs (\mathrm{III})}[\gamma,v] \approx
{\cal E}_{\mathrm{co}}
[\tilde{\tau}_{\mbox{\tiny\textsc{h}e}},v_{\mbox{\tiny\textsc{h}e}}]
_{(\tilde{\tau}_{\mbox{\tiny\textsc{h}e}} = \gamma,\,v_{\mbox{\tiny\textsc{h}e}} = v)}.
\end{equation}
The dominant portion of the correlation energy comes from the electron-repulsion
contribution. The portion arising from the external potential is smaller; since
$v$ is a one-body operator, this term is probably treated well as part of the
first-order energy $E_1$. Therefore, neglecting to make the substitution for the
potential, i.e., $(v_{\mbox{\tiny\textsc{h}e}} = v)$, should only yield a small
error:
\begin{equation} \label{He.approxE} %\mbox{\tiny He.approxE}
E_{\mathrm{co}}^{\scscs (\mathrm{III})}[\gamma] \approx
{\cal E}_{\mathrm{co}}
[\tilde{\tau}_{\mbox{\tiny\textsc{h}e}},v_{\mbox{\tiny\textsc{h}e}}]
_{(\tilde{\tau}_{\mbox{\tiny\textsc{h}e}} = \gamma)}.
\end{equation}
A well known approximation for ${\cal E}_{\mathrm{co}}
[\tilde{\tau}_{\mbox{\tiny\textsc{h}e}},v_{\mbox{\tiny\textsc{h}e}}]$ -- that is
valid for other systems also -- is given by Colle and Salvetti functional
\cite{Colle:75,Lee:88}:
\begin{equation} %\label{} \mbox{\tiny }
{\cal E}_{\mathrm{co}}
[\tilde{\tau}_{\mbox{\tiny\textsc{h}e}},v_{\mbox{\tiny\textsc{h}e}}] \approx
{\cal E}_{\mathrm{co}}^{\mathrm{cs}} [\tilde{\tau}_{\mbox{\tiny\textsc{h}e}}].
\end{equation}
The above two relations suggest the following approximation:
\begin{equation} \label{CS} %\mbox{\tiny CS}
E_{\mathrm{co}}^{\scscs (\mathrm{III})}[\gamma] \approx
{\cal E}_{\mathrm{co}}^{\mathrm{cs}} [\tilde{\tau}_{\mbox{\tiny\textsc{h}e}}]
_{(\tilde{\tau}_{\mbox{\tiny\textsc{h}e}} = \gamma)}.
\end{equation}
While this is probably a reasonable approximation, we do have some reservations
about it. In particular, a universal functional is used to approximate one that
depends on the external potential. This violates, at least, the spirit of the
method. We would therefore prefer to use the approximation given by
Eq.~(\ref{approx7}) or (\ref{He.approxC}). Unfortunately, however, no such
functionals are available. Therefore, let us proceed.

For a closed-shell system, Eq.~(\ref{CS}) becomes
\begin{eqnarray}
\label{CS.v} %\mbox{\tiny CS.v}
\bar{E}_{\mathrm{co}}^{\scscs (\mathrm{III})}[\rho_1]&\approx&
E_{\mathrm{co}}^{\mathrm{cs}}[\rho_1]
=
{\cal E}_{\mathrm{co}}^{\mathrm{cs}} 
[\tilde{\varrho}_{\mbox{\tiny 1\hspace{-0.15ex}\textsc{h}e}}]
_{(\tilde{\varrho}_{\mbox{\tiny 1\hspace{-0.15ex}\textsc{h}e}} = \rho_1)},
\end{eqnarray}
where $\tilde{\varrho}_{\mbox{\tiny1\hspace{-0.15ex}\textsc{h}e}}$ is the
Hartree--Fock spin-less one-particle density-matrix for the helium atom, and, for
future reference, this approximations is denoted by
$E_{\mathrm{co}}^{\mathrm{cs}}[\rho_1]$. An expression for
$E_{\mathrm{co}}^{\mathrm{cs}}[\rho_1]$ is derived by Lee, Parr, and Yang
\cite{Lee:88} that is equivalent to the one derived by Colle--Salvetti; it has the
following form:
\begin{eqnarray} \label{CS.form} %\mbox{\tiny CS.form}
E_{\mathrm{co}}^{\mathrm{cs}}[\rho_1]=
\int G_{\mathrm{co}}
\left(\makebox(0,11){}
\rho_1(\mathbf{r}_1,\mathbf{r}_1),
\nabla_1\rho_1(\mathbf{r}_1,\mathbf{r}_1),
\nabla_1^2\rho_1(\mathbf{r}_1,\mathbf{r}_1),
\left(\makebox(0,10){}\nabla_1\nabla_2
\rho_1(\mathbf{r}_2,\mathbf{r}_1)\right)_{\mathbf{r}_2=\mathbf{r}_1}
\right)
\, d\mathbf{r}_1, 
\end{eqnarray} 
and using this Eq.\ and the previous one, we have
\begin{eqnarray}
\!\!
\bar{E}_{\mathrm{co}}^{\scscs (\mathrm{III})}[\rho_1]&\approx&
\int \!\!\int \delta(\mathbf{r}_2-\mathbf{r}_1)
G_{\mathrm{co}}
\left(\makebox(0,11){}
\rho_1(\mathbf{r}_1,\mathbf{r}_1),
\nabla_1\rho_1(\mathbf{r}_1,\mathbf{r}_1),
\nabla_1^2\rho_1(\mathbf{r}_1,\mathbf{r}_1),
\nabla_1\nabla_2
\rho_1(\mathbf{r}_2,\mathbf{r}_1)\right)
\, d\mathbf{r}_1 \, d\mathbf{r}_2.
\nonumber \\
\end{eqnarray}
When the functional derivative is taken to determine the kernel of the correlation
potential $\nu_{\mathrm{co}}^{\scscs
\rho_1\!\mathrm{III}}(\mathbf{r}_1,\mathbf{r}_2)$ -- as defined by
Eq.~(\ref{Eco.funder}) -- and $\bar{E}_{\mathrm{co}}^{\scscs (\mathrm{III})}$ is
approximated by the above relation, the $\delta(\mathbf{r}_2-\mathbf{r}_1)$ term
survives; the correlation potential $\hat{\nu}_{\mathrm{co}}^{\scscs
\rho_1\!\mathrm{III}}$ is a local operator. Denoting this approximation by
$\nu_{\mathrm{co}}^{\scscs \mathrm{CS}} (\mathbf{r})$, we have
\begin{eqnarray}
\hat{\nu}_{\mathrm{co}}^{\scscs \rho_1\!\mathrm{III}}(\mathbf{r}) \approx
\nu_{\mathrm{co}}^{\scscs \mathrm{CS}} (\mathbf{r}).
\end{eqnarray}

Using the exchange potential given by Eq.~(\ref{B88.pot}) -- or any other local
exchange potential -- and the above local correlation operator, the
exchange-correlation potential $\hat{\nu}_{\mathrm{xc}}^{\scscs \rho_1\!\eta}$ --
given by Eq.~(\ref{operators}) -- is also local. Therefore, the reference-state
$|\Phi\rangle$ obtained from the following non-interacting Hamiltonian:
\begin{eqnarray}
H_s^{\scscs \varrho_1}=
\sum_i
\left(-\mbox{\small$\frac{1}{2}$}
\nabla_{\mbox{\tiny $i$}}^2\right)
+ 
\sum_i
\nu_{s}^{\scscs \varrho_1}(\mathbf{r}_i),
\end{eqnarray}
is also obtained from a local potential, where
\begin{eqnarray}
\nu_{s}^{\scscs \varrho_1}(\mathbf{r}_1)
= v(\mathbf{r}_1) +
\int 
r_{12}^{-1}
\varrho_1(\mathbf{r}_2) 
\,d\mathbf{r}_2
+ v_{\mathrm{x}}^{\scscs \mathrm{ug}}(\mathbf{r}) 
+ v_{\mathrm{x}}^{\scscs \mathrm{B88}}(\mathbf{r}_1)
+ v_{\mathrm{co}}^{\scscs \mathrm{CS}}(\mathbf{r}_1),
\end{eqnarray}
and Eq.~(\ref{eF.canb.sf}), for the occupied orbitals, becomes
\begin{eqnarray} \label{eF.canb.sfB} %\mbox{\tiny eF.canb.sfB}
\left(
-\mbox{\small$\frac{1}{2}$}
\nabla_{\mbox{\tiny $1$}}^2
+ 
\nu_{s}^{\scscs \varrho_1}(\mathbf{r}_1)
\right) 
\chi^{\varrho_1}_w(\mathbf{r}_1)
=
\xi_{w}^{\mbox{\tiny $\varrho_1$}}
\chi^{\varrho_1}_w(\mathbf{r}_1).
\end{eqnarray}
If these equations are solved in a self consistent (SCF) manner, a series of
spin-less, one-particle density-matrices are generated: $\rho_1^0$, $\rho_1^1$,
$\rho_1^2$, $\cdots$; for the $n$th iteration, we have
\begin{eqnarray} \label{eF.canb.scf} %\mbox{\tiny eF.canb.scf}
\left(
-\mbox{\small$\frac{1}{2}$}
\nabla_{\mbox{\tiny $1$}}^2
+ 
\nu_{s}^{\scscs \rho_1^n}(\mathbf{r}_1)
\right) 
\chi^{\rho_1^{n+1}}_w(\mathbf{r}_1)
=
\xi_{w}^{\mbox{\tiny $\rho_1^{n+1}$}}
\chi^{\rho_1^{n+1}}_w(\mathbf{r}_1),
\end{eqnarray}
and this iteration yields a non-interacting state, say $|\Phi^{n+1}\rangle$, with a
one-particle density-matrix of $\rho_1^{n+1}$; in addition, $|\Phi^{n+1}\rangle$
is an eigenfunction of a non-interaction Hamiltonian, $H_s^{\scscs
\rho_1^n}$. From the Hohenberg and Kohn theorem \cite{Hohenberg:64a}, each
non-interacting states generated in the SCF procedure are unique functions of
their electron densities. e.g., $|\Phi^n(\rho)\rangle$; furthermore, because of
the one-to-one correspondence between one-particle density-matrices and
non-interacting states, all one-particle density-matrices generated from the SCF
procedure are determined by their density, i.e., $\mbox{\large
$\rho_1$}(\rho)$. And since for an SCF computation, these are the only
one-particle density-matrices that are required, we can write
\begin{eqnarray} \label{CS.denfun} %\mbox{\tiny CS.denfun}
E_{\mathrm{co}}^{\mathrm{cs}}[\rho_1]=E_{\mathrm{co}}^{\mathrm{cs}}[\rho].
\end{eqnarray}
Because of the general form of $E_{\mathrm{co}}^{\mathrm{cs}}[\rho_1]$ given by
Eq.~(\ref{CS.form}) we need only an expression for $(\nabla_1\nabla_2
\rho_1(\mathbf{r}_2,\mathbf{r}_1))_{\mathbf{r}_2=\mathbf{r}_1}$. Following Lee,
Parr, and Yang, we use the following approximation:
\begin{eqnarray} \label{CS.density} %\mbox{\tiny CS.density}
\left(\makebox(0,11){}\nabla_1\nabla_2
\rho_1(\mathbf{r}_2,\mathbf{r}_1)\right)_{\mathbf{r}_2=\mathbf{r}_1}
\approx \frac{3}{5}(3\pi^2)^{2/3}\rho(\mathbf{r}_1)^{5/3} + \frac{1}{36} 
\frac{|\nabla_1\rho(\mathbf{r}_1)|^2}{\rho(\mathbf{r}_1)} + 
\frac{1}{4} \nabla_1^2 \rho(\mathbf{r}_1),
\end{eqnarray}
and we have
\begin{eqnarray} \label{CS.LYP} %\mbox{\tiny CS.LYP}
E_{\mathrm{co}}^{\mathrm{cs}}[\rho] \approx
E_{\mathrm{co}}^{\mathrm{lyp}}[\rho],
\end{eqnarray}
where $E_{\mathrm{co}}^{\mathrm{lyp}}[\rho]$ is the LYP functional obtained by
substituting the approximation given by Eq.~(\ref{CS.density}) into the
$E_{\mathrm{co}}^{\mathrm{cs}}[\rho_1]$ functional. Using the general of form of
$E_{\mathrm{co}}^{\mathrm{cs}}[\rho_1]$ given by Eq.~(\ref{CS.form}), we have 
\begin{eqnarray}%\label{LYP.denfun} \mbox{\tiny LYP.denfun}
E_{\mathrm{co}}^{\mathrm{lyp}}[\rho]
=
\int G_{\mathrm{co}}
\left(\makebox(0,11){}
\rho(\mathbf{r}),
\nabla\rho(\mathbf{r}),
\nabla^2\rho(\mathbf{r})\right) \, d\mathbf{r}.
\end{eqnarray} 
Using integration by parts, Miehlich, Savin, Stoll, and Preuss, eliminate the
Laplacian terms and obtain the general form given by \cite{Miehloch:89}
\begin{eqnarray}\label{LYP.form} %\mbox{\tiny LYP.form}
E_{\mathrm{co}}^{\mathrm{lyp}}[\rho]
=
\int G_{\mathrm{co}}
\left(\makebox(0,11){}
\rho(\mathbf{r}),
\nabla\rho(\mathbf{r})\right) \, d\mathbf{r}.
\end{eqnarray} 
Using Eqs.~(\ref{CS.v}), (\ref{CS.denfun}), (\ref{CS.LYP}), (\ref{LYP.form}), and
(\ref{ident.grad}) we obtain the following approximation for the
correlation-energy functional:
\begin{eqnarray} \label{lyp2} %\mbox{\tiny lyp2}
\bar{E}_{\mathrm{co}}^{\scscs (\mathrm{III})}[\rho_1]&\approx&
E_{\mathrm{co}}^{\mathrm{lyp}}[\rho]
=
\int \!\!\int \delta(\mathbf{r}_2-\mathbf{r}_1)
G_{\mathrm{co}}
\left(\makebox(0,11){}
\rho_1(\mathbf{r}_2,\mathbf{r}_1),
(\nabla_1 + \nabla_2)\rho_1(\mathbf{r}_2,\mathbf{r}_1)
\right)
\, d\mathbf{r}_1 \, d\mathbf{r}_2, \nonumber \\
\end{eqnarray}
and using the results from Appendix \ref{pot.ident}, the local correlation
potential is given by
\begin{eqnarray}
\hat{\nu}_{\mathrm{co}}^{\scscs \rho_1\!\mathrm{III}}(\mathbf{r}) \approx
\nu_{\mathrm{co}}^{\scscs \mathrm{LYP}} (\mathbf{r}) = 
\frac{\delta E_{\mathrm{co}}^{\mathrm{lyp}}[\rho]} 
{\delta \rho(\mathbf{r})}.
\end{eqnarray}
Using these two approximations for $E_{\mathrm{co}}^{\mathrm{lyp}}[\rho]$ and
$\hat{\nu}_{\mathrm{co}}^{\scscs \rho_1\!\mathrm{III}}(\mathbf{r})$ yields an
identical approach as the LYP method. If, in addition, the Becke exchange
functional, given by Eq.~(\ref{b88}), is used, the method is equivalent to the one
known as BLYP.

\subsection{The B3LYP Functional}

Using Eq.~(\ref{exco.sfv}), and the uniform-electron-gas approximation for the
correlation-energy functional, Eq.~(\ref{ldaF}), we obtain an approximate
exchange-correlation functional, given by
\begin{eqnarray} %\label{} \mbox{\tiny }
\bar{E}_{\mathrm{xc}}^{\scscs (\mathrm{III})}[\rho_1]
\approx
E_{\mathrm{x}}[\rho_1] +
E_{\mathrm{co}}^{\mathrm{ug}}[\rho]. 
\end{eqnarray} 
An alternative approximation is given by the BLYP functional that uses Eqs.~(\ref{b88})
and (\ref{lyp2}):
\begin{eqnarray} %\label{} \mbox{\tiny }
\bar{E}_{\mathrm{xc}}^{\scscs (\mathrm{III})}[\rho_1]
\approx
E_{\mathrm{x}}^{\mathrm{ug}}[\rho] + \Delta
\!E_{\mathrm{x}}^{\scscs \mathrm{B88}}[\rho]+
E_{\mathrm{co}}^{\mathrm{lyp}}[\rho].
\end{eqnarray} 
If both of these approximations are reasonable, then so is a linear combination of
of the two:
\begin{eqnarray} %\label{} \mbox{\tiny }
\bar{E}_{\mathrm{xc}}^{\scscs (\mathrm{III})}[\rho_1]
\approx
b\left(E_{\mathrm{x}}^{\mathrm{ug}}[\rho] + \Delta
\!E_{\mathrm{x}}^{\scscs \mathrm{B88}}[\rho] +
E_{\mathrm{co}}^{\mathrm{lyp}}[\rho]\right)
+
(1-b)\left(E_{\mathrm{x}}[\rho_1] + E_{\mathrm{co}}^{\mathrm{ug}}[\rho]\right),
\end{eqnarray} 
where the semi-empirical parameter $b$ is at our disposal. Using the approximation
for $E_{\mathrm{co}}^{\mathrm{ug}}$, given by Vosko, Wilk, and Nusair
\cite{Vosko:80}, denoted by $E_{\mathrm{co}}^{\mathrm{vwn}}$, we have
\begin{eqnarray} %\label{} \mbox{\tiny }
\bar{E}_{\mathrm{xc}}^{\scscs (\mathrm{III})}[\rho_1]
\approx
b\left(E_{\mathrm{x}}^{\mathrm{ug}}[\rho] + \Delta
\!E_{\mathrm{x}}^{\scscs \mathrm{B88}}[\rho]
+ E_{\mathrm{co}}^{\mathrm{lyp}}[\rho]\right)
+
(1-b)\left(E_{\mathrm{x}}[\rho_1] + E_{\mathrm{co}}^{\mathrm{vwn}}[\rho]\right).
\end{eqnarray} 
Insertion of additional empirical parameters that deviate slightly from
unity, we have
\begin{eqnarray} \label{b3lyp} %\mbox{\tiny b3lyp} 
\bar{E}_{\mathrm{xc}}^{\scscs (\mathrm{III})}[\rho_1]
\approx
b\left(E_{\mathrm{x}}^{\mathrm{ug}}[\rho] + \alpha \Delta
\!E_{\mathrm{x}}^{\scscs \mathrm{B88}}[\rho]
+ \beta E_{\mathrm{co}}^{\mathrm{lyp}}[\rho]\right)
+
(1-b)\left(E_{\mathrm{x}}[\rho_1]+ \gamma E_{\mathrm{co}}^{\mathrm{vwn}}[\rho] \right),
\end{eqnarray}
where we can reduce the four variables to three by introducing a constraint, e.g.,
($\gamma[b,\beta]$). By choosing ($b=0.80$), ($\alpha=0.90$), ($\beta=1.01$),
($\gamma=(1-\beta b)(1-b)^{-1}=0.95$) we get the same exchange-correlation energy
functional as the one employed in the B3LYP approach. In particular, we have
($b=1-a_0$), ($b\alpha=a_x$), and ($b\beta=a_c$), where $a_0$, $a_x$, and $a_c$,
are 0.20, 0.72, and 0.81, respectively. 

\section{Acknowledgments}
The author thanks Kimihiko Hirao for useful discussions.

\appendix 

\section{Derivation of some identities} \label{pot.ident}

Consider the following density functional: 
\begin{eqnarray} %\label{} \mbox{\tiny }
E_{i}[\rho]=
\int G_i(\rho)
\,d\mathbf{r},
\end{eqnarray}
and a corresponding potential, given by \cite{Gelfand:63,Parr:89,Handy:97}
\begin{eqnarray} \label{poten} %\mbox{\tiny poten}
v_{i}(\mathbf{r}) = \frac{\delta E_{i}[\rho]}{\delta\rho(\mathbf{r})}
= \frac{\partial G_i(\rho)}{\partial \rho(\mathbf{r})}.
\end{eqnarray}
If the integrand $G_i(\rho)$ does not contain gradient terms, we can express the
density functional as one that appears to depends on the spin-less one-particle
density-matrix:
\begin{equation} %\label{} \mbox{\tiny}
E_{i}[\rho]=E_{i}[\rho_1]=
\int \!\! \int \delta(\mathbf{r}_1-\mathbf{r}_2) G_i(\rho_1)
\,d\mathbf{r}_1
\,d\mathbf{r}_2,
\end{equation}
and its corresponding operator, denoted by $\hat{v}_{i}$, is defined by its
kernel, given by
\begin{eqnarray} \label{kern} %\mbox{\tiny kern}
v_{i}(\mathbf{r}_1,\mathbf{r}_2) = 
\frac{\delta E_{i}[\rho_1]}{\delta\rho_1(\mathbf{r}_2,\mathbf{r}_1)}
= \delta(\mathbf{r}_1-\mathbf{r}_2)
\frac{\partial G_i(\rho_1)}{\partial \rho_1(\mathbf{r}_2,\mathbf{r}_1)}.
\end{eqnarray}
Operating with $\hat{v}_{i}$ upon an arbitrary function, $\phi$, and using
Eqs.~(\ref{poten}) and (\ref{kern}), we obtain the following:
\begin{eqnarray} %\label{} \mbox{\tiny }
\hat{v}_{i} \phi(\mathbf{r}_1) =
\int v_{i}(\mathbf{r}_1,\mathbf{r}_2)\phi(\mathbf{r}_2) \,d\mathbf{r}_2
= \int \delta(\mathbf{r}_1-\mathbf{r}_2)
\frac{\partial G_i\left(\rho_1(\mathbf{r}_2,\mathbf{r}_1)\right)}
{\partial \rho_1(\mathbf{r}_2,\mathbf{r}_1)}
\phi(\mathbf{r}_2) \,d\mathbf{r}_2 
\nonumber \\
= \frac{\partial G_i\left(\rho_1(\mathbf{r}_1,\mathbf{r}_1)\right)}
{\partial \rho_1(\mathbf{r}_1,\mathbf{r}_1)}
\phi(\mathbf{r}_1)=
\frac{\partial G_i\left(\rho(\mathbf{r}_1)\right)}
{\partial \rho(\mathbf{r}_1)}
\phi(\mathbf{r}_1)=
v_{i}(\mathbf{r}_1)\phi(\mathbf{r}_1).
\end{eqnarray}
Hence, the potential and operator are identical:
\begin{eqnarray} %\label{} \mbox{\tiny }
\hat{v}_{i}=v_{i}(\mathbf{r}).
\end{eqnarray}

Now let the energy functional have a gradient dependence:
\begin{eqnarray} %\label{} \mbox{\tiny }
E_{i}[\rho]=
\int G_i(\rho,\nabla\rho)
\,d\mathbf{r},
\end{eqnarray}
where its corresponding potential is given by \cite{Gelfand:63,Parr:89,Handy:97}
\begin{eqnarray} \label{poten2} %\mbox{\tiny poten2}
v_{i}(\mathbf{r}) = \frac{\delta E_{i}[\rho]}{\delta\rho(\mathbf{r})}
= \frac{\partial G_i(\rho)}{\partial \rho(\mathbf{r})}
- \frac{d}{d\mathbf{r}}\cdot
\frac{\partial G_i(\rho)}{\partial \nabla\rho(\mathbf{r})}.
\end{eqnarray}
Using the identity~(\ref{ident.grad}), we can express the density functional as a
one-particle density-matrix functional:
\begin{equation} %\label{} \mbox{\tiny}
E_{i}[\rho]=E_{i}[\rho_1]=
\int \!\! \int \delta(\mathbf{r}_1-\mathbf{r}_2) 
G_i\left(\rho_1(\mathbf{r}_2,\mathbf{r}_1),
\left(\mbox{\small $\nabla_1$}+\mbox{\small $\nabla_2$}\right)
\rho_1(\mathbf{r}_2,\mathbf{r}_1)
\makebox(0,11){}
\right)
\,d\mathbf{r}_1
\,d\mathbf{r}_2,
\end{equation}
and its corresponding operator, denoted by $\hat{v}_{i}$, is defined by its
kernel, given by
\begin{eqnarray} \label{kern2} %\mbox{\tiny kern2}
v_{i}(\mathbf{r}_1,\mathbf{r}_2)&=& 
\frac{\delta E_{i}[\rho_1]}{\delta\rho(\mathbf{r}_2,\mathbf{r}_1)}
\\
&=& 
\delta(\mathbf{r}_1-\mathbf{r}_2)
\left(
\frac{\partial G_i(\rho_1)}{\partial \rho_1(\mathbf{r}_2,\mathbf{r}_1)}
- \frac{d}{d\mathbf{r}_1}\cdot
\frac{\partial G_i(\rho_1)}{\partial \nabla_1\rho_1(\mathbf{r}_2,\mathbf{r}_1)}
- \frac{d}{d\mathbf{r}_2}\cdot
\frac{\partial G_i(\rho_1)}{\partial \nabla_2\rho_1(\mathbf{r}_2,\mathbf{r}_1)}
\right). \nonumber 
\end{eqnarray}
Operating with $\hat{v}_{i}$ upon an arbitrary function $\phi$, using
Eqs.~(\ref{kern2}) and (\ref{poten2}), and an identity, given by
\begin{eqnarray} %\label{} \mbox{\tiny }
\left(\frac{d}{d\mathbf{r}_1}\cdot
\frac{\partial G_i(\rho_1)}{\partial \nabla_1\rho_1(\mathbf{r}_2,\mathbf{r}_1)}
+ \frac{d}{d\mathbf{r}_2}\cdot
\frac{\partial G_i(\rho_1)}{\partial \nabla_2\rho_1(\mathbf{r}_2,\mathbf{r}_1)}
\right)_{\mathbf{r}_2=\mathbf{r}_1}
=
\frac{d}{d\mathbf{r}_1}\cdot
\frac{\partial G_i(\rho)}{\partial \nabla_1\rho(\mathbf{r}_1)},
\end{eqnarray}
we obtain the following:
\begin{eqnarray} %\label{} \mbox{\tiny }
\nonumber 
\hat{v}_{i} \phi(\mathbf{r}_1)&=&
\int v_{i}(\mathbf{r}_1,\mathbf{r}_2)\phi(\mathbf{r}_2) \,d\mathbf{r}_2\
\\ &=&
\int \delta(\mathbf{r}_1-\mathbf{r}_2)
\left(
\frac{\partial G_i(\rho_1)}{\partial \rho_1(\mathbf{r}_2,\mathbf{r}_1)}
- \frac{d}{d\mathbf{r}_1}\cdot
\frac{\partial G_i(\rho_1)}{\partial \nabla_1\rho_1(\mathbf{r}_2,\mathbf{r}_1)}
- \frac{d}{d\mathbf{r}_2}\cdot
\frac{\partial G_i(\rho_1)}{\partial \nabla_2\rho_1(\mathbf{r}_2,\mathbf{r}_1)}
\right)
\phi(\mathbf{r}_2) \,d\mathbf{r}_2 
\nonumber \\ \nonumber \\
&=& \left[
\frac{\partial G_i\left(\rho(\mathbf{r}_1)\right)}
{\partial \rho(\mathbf{r}_1)} 
- \left(\frac{d}{d\mathbf{r}_1}\cdot
\frac{\partial G_i(\rho_1)}{\partial \nabla_1\rho_1(\mathbf{r}_2,\mathbf{r}_1)}
+ \frac{d}{d\mathbf{r}_2}\cdot
\frac{\partial G_i(\rho_1)}{\partial \nabla_2\rho_1(\mathbf{r}_2,\mathbf{r}_1)}
\right)_{\mathbf{r}_2=\mathbf{r}_1}
\right]
\phi(\mathbf{r}_1) \nonumber
\\ \nonumber \\
&=&\left(
\frac{\partial G_i\left(\rho(\mathbf{r}_1)\right)}
{\partial \rho(\mathbf{r}_1)} 
- \frac{d}{d\mathbf{r}_1}\cdot
\frac{\partial G_i(\rho)}{\partial \nabla_1\rho(\mathbf{r}_1)}
\right)
\phi(\mathbf{r}_1)=
v_{i}(\mathbf{r}_1)\phi(\mathbf{r}_1).
\end{eqnarray}
Hence, we have
\begin{eqnarray} %\label{} \mbox{\tiny }
\hat{v}_{i}=v_{i}(\mathbf{r}).
\end{eqnarray}

\bibliography{ref}
\end{document}